\documentclass[12pt,epsf,qsymbols]{article}
\pdfoutput=1
\usepackage[utf8]{inputenc}
\usepackage[normalem]{ulem}
\usepackage[english]{babel}
\usepackage{tabularx}
\usepackage{array}
\usepackage{graphics}
\usepackage[pdftex]{graphicx}
\usepackage{psfrag}
\usepackage{epsfig}
\usepackage{subfig}
\usepackage{amsmath}
\usepackage{mathtools}
\usepackage{amssymb}
\usepackage{bbold}
\usepackage{ulem}
\usepackage{setspace}
\usepackage{rotating}
\usepackage{colortbl}
\usepackage{longtable}
\usepackage{slashed}
\usepackage{braket}
\usepackage{lineno}
\makeatletter
\usepackage{textcomp}
\usepackage[usenames,dvipsnames]{xcolor}
\usepackage{relsize}
\usepackage{cite}
\usepackage{cancel}
\usepackage{ulem}

\usepackage{float}

\usepackage{bbm}
\usepackage{bm}
\usepackage{enumerate}
\usepackage{amsmath}
\usepackage{verbatim}

\usepackage{fancyhdr}

\usepackage{multirow}
\usepackage{arydshln}
\usepackage{enumitem}

\usepackage{hyperref}
\hypersetup{colorlinks,citecolor=nicegreen,linkcolor=niceblue}
\hypersetup{colorlinks=true}

\usepackage{pifont}

\allowdisplaybreaks

\setlongtables

\newcommand{\bea}{\begin{eqnarray}}
\newcommand{\eea}{\end{eqnarray}}
\newcommand{\beq}{\begin{equation}}
\newcommand{\eeq}{\end{equation}}
\newcommand{\ec}{\end{center}}
\newcommand{\bc}{\begin{center}}

\newcommand{\pdir}{p\kern -5.2pt\raise 0.2ex\hbox {/}}

\newcommand{\vdir}{v\kern -5.75pt\raise 0.15ex\hbox {/}}
\newcommand{\kdir}{k\kern -5.75pt\raise 0.15ex\hbox {/}}
\newcommand{\epsdir}{\epsilon\kern -5.0pt\raise 0.15ex\hbox {/}}
\newcommand{\bvdir}{\bar{v}\kern -5.75pt\raise 0.15ex\hbox {/}}
\newcommand{\Ddir}{D\kern -7.75pt\raise 0.20ex\hbox {/}}
\newcommand{\Adir}{A\kern -7.75pt\raise 0.20ex\hbox {/}}
\newcommand{\ldir}{l\kern -5.0pt\raise 0.2ex\hbox{/}}
\newcommand{\varepsdir}{\varepsilon\kern -5.5pt\raise 0.15ex\hbox{/}}

\newcommand{\abs}[1]{\left| #1 \right|}

\newcommand{ \mysmall}[1]{\scriptscriptstyle #1} 

\makeatother

\definecolor{niceblue}{rgb}{0.15,0.15,0.6}
\definecolor{nicegreen}{rgb}{0.1,0.5,0.1}
\definecolor{Red}{rgb}{1.,0.,0.}

\definecolor{Green}{rgb}{0.2,.7,0.2}

\begin{document}
\unitlength = 1mm

\thispagestyle{empty} 

\begin{flushright}
\begin{tabular}{l}
{\tt \footnotesize \color{blue}ZU-TH 46/19}\\ 
\end{tabular}
\end{flushright}
\begin{center}
\vskip 3.4cm\par
{\par\centering \textbf{\Large   \bf Hunting for ALPs with Lepton Flavor Violation}}
\vskip 1.2cm\par
{\scalebox{.85}{\par\centering \large  
\sc Claudia~Cornella$^{a,b}$, Paride~Paradisi$^{b}$, and Olcyr~Sumensari$^{b}$}
{\par\centering \vskip 0.7 cm\par}
{\sl 
$^a$~{Physik-Institut, Universität Zürich, CH-8057 Zürich, Switzerland}}\\
{\par\centering \vskip 0.25 cm\par}
{\sl 
$^b$~Istituto Nazionale Fisica Nucleare, Sezione di Padova, I-35131 Padova, Italy
Dipartimento di Fisica e Astronomia ``G.~Galilei", Universit\`a degli Studi di Padova, Italy}\\

{\vskip 1.65cm\par}}

\end{center}

\vskip 0.85cm
\begin{abstract}
We examine the low-energy signatures of axion-like particles (ALPs) in lepton flavor violating (LFV) processes. By using a dimension-5 effective Lagrangian, we compute the most general ALP contributions to LFV decays of leptons and mesons. The provided expressions are valid for any choice of ALP mass and couplings. We explore the complementarity of different processes, identifying specific patterns to be experimentally tested.
Constraints on LFV couplings are derived from existing data and prospects for forthcoming experiments are also discussed. As a by-product, we revisit the possibility of a simultaneous explanation of the observed discrepancies in the muon and electron $g-2$ through ALP interactions.
\end{abstract}
\newpage
\setcounter{page}{1}
\setcounter{footnote}{0}
\setcounter{equation}{0}
\noindent

\renewcommand{\thefootnote}{\arabic{footnote}}

\setcounter{footnote}{0}

\tableofcontents

\newpage

\section{Introduction}
\label{sec:intro}

After the long-awaited discovery of the Higgs boson at the Large Hadron Collider (LHC), no convincing evidence of heavy new physics (NP) has emerged yet. The motivation for NP at the TeV scale has thus being questioned, and alternative scenarios, with new light mediators, 
have received increasing attention from both experimental and theoretical communities over the past several years.
A prominent example are Standard Model (SM) extensions with light pseudoscalar bosons, generically referred to as axion-like-particles (ALPs)~\cite{Jaeckel:2010ni}. ALPs might be naturally light, in comparison to the NP scale they originate from, if they are 
pseudo-Nambu-Goldstone-Bosons (pNGBs) of an underlying broken symmetry. Such particles can be motivated by a number of fundamental 
open questions in particle physics such as the strong CP problem~\cite{Peccei:1977hh}, the origin of dark matter~\cite{Preskill:1982cy}, 
as well the hierarchy~\cite{Dvali:2003br} and flavor problems~\cite{Wilczek:1982rv}. 

Within specific ultraviolet scenarios, the ALP mass and couplings are typically related. Alternatively, a more model independent approach is to consider ALPs as a generalization of the QCD axion, with mass and couplings being free parameters to be probed experimentally. In this context, ALP interactions with SM fermions and gauge bosons are described via an effective Lagrangian built with operators up to dimension-5~\cite{Georgi:1986df}, and they can be probed in a wide range of experimental facilities.  Although less predictive, this approach has the advantage of capturing general features of a broad class of models. 

The most stringent limits on ALP couplings arise from cosmological and astrophysical bounds, which are valid for ALP masses below the MeV 
scale~\cite{Jaeckel:2010ni,DiLuzio:2016sbl}. Less severe but still significant bounds are set by LEP for masses ranging from the MeV scale up to $90$~GeV~\cite{Jaeckel:2015jla}. The possibility of probing ALPs at the LHC, as well as in future colliders and 
fixed-target facilities, has also been extensively explored~\cite{Knapen:2016moh,Bauer:2017ris}. Furthermore, the rich research program at flavor factories offers several opportunities to probe ALP parameters in meson decays~\cite{Dolan:2014ska}. In particular, since there is no fundamental reason for the ALP interactions to respect the SM flavor group, ALPs can induce flavor-changing neutral-currents (FCNC) already at tree-level~\cite{Batell:2009jf,Gavela:2019wzg}.

The main goal of this paper is to explore the discovery potential of ALPs through lepton flavor violating (LFV) processes. Since lepton flavor is an accidental symmetry of the SM, broken by the tiny neutrino masses, its observation would correspond to an unambiguous manifestation of new physics. This is an extremely promising and timely subject thanks to the ongoing experimental program at present NA62~\cite{Petrov:2017wza}, LHCb~\cite{Bediaga:2018lhg} and Belle-II~\cite{Kou:2018nap}, as well as at the future Mu2E~\cite{Bartoszek:2014mya}, Mu3E~\cite{Blondel:2013ia} and COMET~\cite{Adamov:2018vin} experiments, which will improve the current sensitivity by orders of magnitude, see~Tables~\ref{tab:exp-lep} and \ref{tab:exp}. The main focus in the literature so far has been the on-shell ALP contribution to processes such as $\mu\to ea \to eee$ and $\mu\to ea \to e\gamma\gamma$, due to the resonantly enhanced decay rates~\cite{Heeck:2017xmg}. For ALP masses above the kinematical threshold of on-shell production, i.e.~$m_a> m_\mu-m_e$, there is a nontrivial interplay with loop-induced processes such as $\mu\to e \gamma$ which has not been fully explored yet. Furthermore, LFV meson decays can provide complementary limits to the ones derived from purely leptonic processes, as they are sensitive to different combinations of ALP couplings.

In this paper, we provide a comprehensive study of ALP-induced LFV decays, generalizing and complementing previous analyses~\cite{Heeck:2017xmg,Bauer:2019gfk}. We consider two distinct classes of processes: (i) purely leptonic and (ii) hadronic decays. The first class comprises radiative decays $\ell_j \to \ell_i \gamma$, three-body 
decays $\ell_j \to \ell_i \ell_k\ell_k$ and $\ell_j \to \ell_i \gamma\gamma$ (with $\ell_{j}= \mu,\tau$), as well as $\mu\to e$ conversion in nuclei, which we consider to be a leptonic process since the only coherently-enhanced contributions are the ones coming from photon penguins. Hadronic decays include leptonic meson decays $P\to \ell_i\ell_j$, their inverse processes $\tau \to P \ell_j$, 
as well as semileptonic decays $P\to P^\prime (V)\ell_i\ell_j$, where $P^{(\prime)}$ and $V$ stand for pseudoscalar and vector mesons, respectively. We derive general formulae for these processes that can be applied for any choice of the ALP mass and couplings. We show in detail that the relative importance of the various processes depends not only on the relative strength of ALP couplings to photon/gluons and fermions but also on the ALP mass. In particular, the correlations among observables can change drastically depending on whether ALPs are produced on-shell or off-shell. This makes it possible to infer the ALP mass indirectly, through their virtual effects to LFV processes.

Another observable which is highly sensitive to ALP couplings is the anomalous magnetic moment of leptons $a_{\ell} = (g-2)_\ell/2$~\cite{Marciano:2016yhf,Bauer:2017ris}. This quantity received a lot of attention due to the longstanding $\approx 3.6\sigma$ discrepancy between the experimental measurement and the SM prediction, $\Delta a_\mu= a^{\mysmall\rm exp}_{\mu} - a^{\mysmall\rm SM}_{\mu} = (27.1\pm 7.3) \times 10^{-10}$ \cite{Bennett:2006fi,Keshavarzi:2018mgv,Davier:2017zfy}, which might be soon clarified by the ongoing analysis at the Muon g-2 experiment (Fermilab)~\cite{Grange:2015fou}. In a large class of NP scenarios, contributions to magnetic moment scale with the square of lepton masses (the so-called ``naive scaling''), in such a way that the current discrepancy $\Delta a_\mu$ suggests a NP effect in $a_e$ at the level of $(7 \pm 2) \times 10^{-14}$~\cite{Giudice:2012ms}.~\footnote{See Ref.~\cite{Giudice:2012ms,Davoudiasl:2018fbb} for examples of new physics scenarios that violate the ``naive scaling" rule, thus allowing for larger effects in $\Delta a_e$ than in $\Delta a_\mu$.} Testing NP with $(g-2)_{e}$ became possible only very recently thanks to the improved measurement of the fine-structure constant $\alpha_{\mathrm{em}}$ from atomic physics experiments~\cite{Parker:2018vye}. Remarkably, the reevaluation of $\Delta a_e$ employing the latest $\alpha_\mathrm{em}$ value has shown a $2.4\sigma$ deviation from the SM prediction, namely $\Delta a_e=(-87\pm 36)\times 10^{-14}$~\cite{Hanneke:2008tm}, which departs considerably from the ``naive scaling" expectation and  has the opposite sign compared to $\Delta a_\mu$. Another goal of this paper is to assess to which extent ALPs can explain simultaneously 
both $g-2$ deviations, exploring the potential interplay with LFV observables (see also Ref.~\cite{Bauer:2019gfk} for a recent discussion). 

\

The paper is organized as follows. In Sec.~\ref{sec:eft}, we recall the effective description of ALP interactions in terms of dimension-five operators. In Sec.~\ref{sec:lep}, we discuss the purely leptonic probes of LFV ALP couplings, deriving general expressions for the relevant observables and extracting constraints from available experimental results. In Sec.~\ref{sec:had}, we extend our discussion to LFV processes involving hadrons. Our findings are then summarized in Sec.~\ref{sec:conclusion}.

\section{Effective field theory description}
\label{sec:eft}

At low energies, the most general dimension-5 effective Lagrangian describing ALP interactions with fermions and photons/gluons reads~\cite{Georgi:1986df}
\begin{align}
\label{eq:leff}
\begin{split}
\mathcal{L}_{\mathrm{eff}}^{\mathrm{d\leq 5}} &= \dfrac{1}{2} (\partial_\mu a) (\partial^\mu a)- \dfrac{m_a^{2} a^2}{2}\\
&+e^2\,c_{\gamma\gamma}\dfrac{a}{\Lambda} F_{\mu\nu} \widetilde{F}^{\mu\nu}
+g_s^2\,c_{gg}\dfrac{a}{\Lambda} G_{\mu\nu} \widetilde{G}^{\mu\nu}-\dfrac{\partial_\mu a}{\Lambda} \sum_{f,i,j} \bar{f}_i \gamma^\mu(v_{ij}^f-a_{ij}^f \gamma_5)f_j\,,
\end{split}
\end{align}
\noindent where the invariance under the gauge symmetry $SU(3)_c \times U(1)_{\mathrm{em}}$ has been imposed. In this equation, 
$f \in \lbrace u,d,\ell \rbrace$ denotes SM fermions in the mass basis, $i,j$ stand for flavor indices and the dual field strengths are defined as $\widetilde{X}^{\mu\nu}=  \frac{1}{2}\varepsilon^{\mu\nu\alpha\beta} X_{\alpha\beta}$, with $\varepsilon^{0123}=+1$. The effective couplings to photons and gluons are denoted by $c_{\gamma\gamma}$ and $c_{gg}$, while $v_{ij}^f$ and $a_{ij}^f$ stand for the vector and axial-vector ALP couplings to SM fermions, and $\Lambda$ for the EFT cutoff.~\footnote{We factor out the gauge couplings in Eq.~\eqref{eq:leff} in such a way that the Wilson coefficients $c_{\gamma\gamma}$ and $c_{gg}$ become scale invariant at one-loop order \cite{Bauer:2017ris}.} The matrices $v^{f}$ and $a^{f}$ are taken to be hermitian and real to avoid CP violating effects. By using the equations of motion, the last term in Eq.~\eqref{eq:leff}
can be recast as
\begin{align}
\label{eq:yuk-alp}
\begin{split}
\mathcal{L}_{\mathrm{eff}}^{\mathrm{d\leq 5}} \supset -i \dfrac{a}{\Lambda} \sum_{f,i,j} &\bar{f}_i \big{[}(m_{f_j}-m_{f_i})\,v_{ij}^f +(m_{f_j}+m_{f_i})\, a_{ij}^f\, \gamma_5\big{]} f_j\,.
\end{split}
\end{align}

\noindent The anomaly equation for the axial-vector current divergence also entails a modification of $c_{\gamma\gamma}$ and $c_{gg}$ couplings, see e.g.~discussion in Ref.~\cite{Bauer:2017ris}. In the following, we will denote the effective couplings accounting for the full one-loop contributions as $c_{\gamma\gamma}\equiv c_{\gamma\gamma}^{\mathrm{eff}}$ and $c_{gg}\equiv c_{gg}^{\mathrm{eff}}$.  From Eq.~\eqref{eq:yuk-alp}, we see that $v_{ij}^f$ contributes to flavor-violating observables only, as expected from the vector-current conservation, while $a^f_{ij}$ enters both flavor-conserving and violating observables. For future convenience, we define the effective couplings
$s_{ij}^\ell$
\begin{equation}
s_{ij}^\ell \equiv \sqrt{|a_{ij}^\ell|^2+|v_{ij}^\ell|^2}\,,
\end{equation}
\noindent which appears in the expressions of most LFV processes, as shown below. 

Before studying the phenomenological implications of the effective Lagrangian defined above, we discuss the theoretical bounds on the ALP couplings arising from perturbative unitarity~\cite{Lee:1977yc}. 
Indeed, for sufficiently large values of the energy $\sqrt{s}$, the EFT description is expected to break down. We estimate these constraints by computing the partial wave unitarity bounds on $\gamma\gamma \to \gamma\gamma$, 
$gg \rightarrow gg$ and $\bar f f \rightarrow \bar f f$ amplitudes mediated by a pseudoscalar boson in the limit of high-energies
($\sqrt{s}\gg m_a$). By requiring partial waves of total angular momentum $J=0$ to satisfy $|{\rm Re}\, a_0|< 1/2$, we obtain the following conditions
\begin{align}
\sqrt{s} &<  \frac{\Lambda}{\sqrt{4\pi} \alpha_{\mathrm{em}} \, |c_{\gamma\gamma}|}\,,      \qquad \qquad\qquad
\sqrt{s} <  \frac{\Lambda}{\sqrt{32\pi} \alpha_{s} \, |c_{gg}|}\,,
\label{eq:bound_unitary_gauge}
\\
 |v_{ij}^f| &< \sqrt{\frac{8\pi}{3}}  \dfrac{\Lambda}{(m_{f_j} - m_{f_i})}\,,         \qquad  \qquad
\hspace*{-0.18em}|a_{ij}^f| < \sqrt{\frac{8\pi}{3}}  \dfrac{\Lambda}{(m_{f_j} + m_{f_i})}\,.
\label{eq:bound_unitary_yukawa}
\end{align}
As an example, if $\Lambda = 1~$TeV and $c_{gg} \, (c_{\gamma\gamma}) = 1$, from Eq.~\eqref{eq:bound_unitary_gauge} we learn that our EFT remains unitary up to energies $\sqrt{s} \lesssim 10(40)~$TeV. 
On the other hand, bounds on the Yukawa couplings $v_{ij}^f$, and $a_{ij}^f$ do not depend on $\sqrt{s}$. 
For instance, for $\Lambda = 1~$TeV, Eq.~\eqref{eq:bound_unitary_yukawa} yields $|v_{3i}^f|,\, |a_{3i}^f| \lesssim 2000$ (with $i<3$).

We now proceed to examine several LFV observables sensitive to the couplings defined above, starting with purely leptonic processes.

\section{Purely leptonic processes}
\label{sec:lep}

The most promising decay channels in this category are the radiative decays $\ell_j \to \ell_i \,\gamma$, and the 
three-body decays $\ell_j\to \ell_i \ell_k \ell_k$ and $\ell_j\to \ell_i \gamma\gamma$ (with $i<j$). In the following, we provide the general 
expressions for each of these processes and discuss their potential in constraining ALP couplings.
\subsection{$\ell_{j} \to \ell_{i} \gamma$ }
\label{ssec:dipole}

\begin{figure}[h!]
\centering
\includegraphics[width=0.65\textwidth]{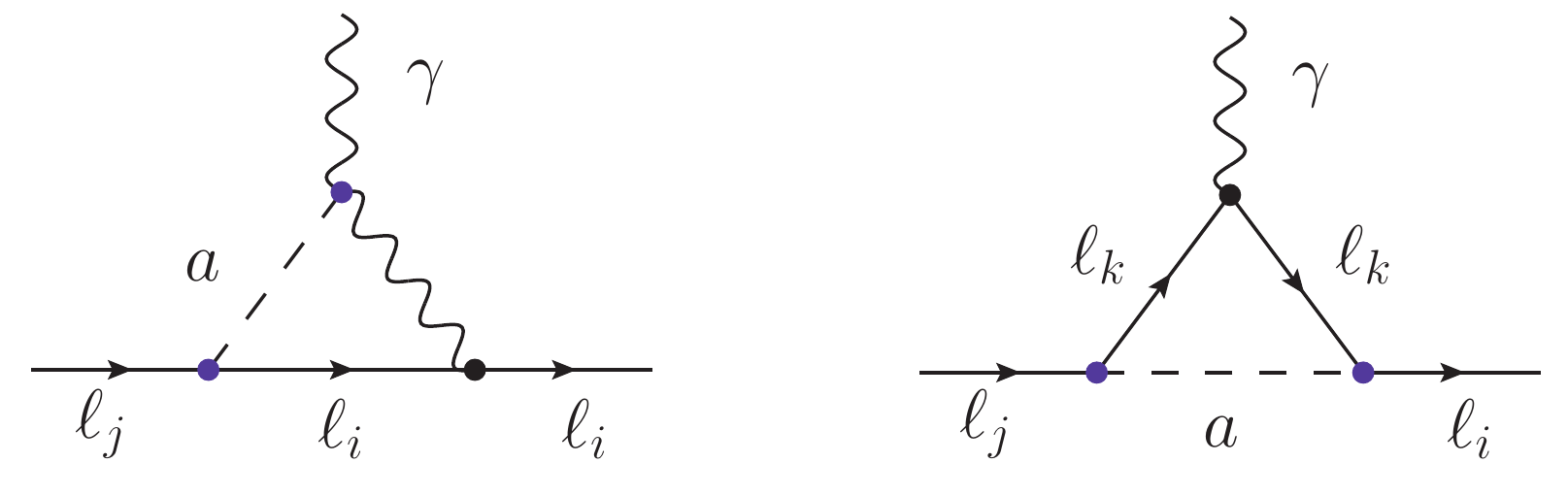}
\caption{\sl \small Diagrams illustrating the ALP contributions to $\ell_j \to \ell_i\gamma$ at linear (left panel) and quadratic (right panel) order in the Yukawa couplings.}
\label{fig:diagram-mueg}
\end{figure}

The amplitude for the process $\ell_{j} \to \ell_{i} \gamma$  can be generically parameterized as

\begin{equation}
\label{eq:leff-dip}
i \mathcal{M}^{\mu}(\ell_j\to\ell_i \gamma)  =  i \, e \, \bar u_{i} (p-q) \Sigma^{\mu}_{ij}(0) u_{j}(p)\,,
\end{equation}

\noindent where $p$ and $q$ denote the momentum of the heavy lepton and the photon, respectively. Since the photon is on-shell in this process, gauge invariance implies that the most general Dirac structure is given by
\begin{equation}
\Sigma_{ij}^\mu(0) = \frac{  i \sigma^{\mu \nu}    q_{\nu} }{m_{\ell_j}} \left(\mathcal F^{ij}_{2} (0) + \mathcal G^{ij}_{2} (0) \gamma^{5}  \right)  \,,
\end{equation}
\noindent where $\mathcal{F}^{ij}_{2} (0)$ and $\mathcal{G}^{ij}_{2} (0)$ are dimensionless form-factors. 
The above expression allows us to write the decay rate as
\begin{align}
\label{eq:Bmueg}
\Gamma(\ell_{j} \to \ell_{i} \gamma) = \frac{m_{\ell_j} }{8 \pi} e^{2} \left( \abs{\mathcal{F}_{2}^{ij}(0)}^{2} +  \abs{\mathcal{G}_{2}^{ij}(0)}^{2} \right) \,,
\end{align}

\noindent where $m_{\ell_j}\gg m_{\ell_i}$ has been used. The leading contributions to the dipole form-factors $\mathcal{F}^{ij}_{2} (0)$ and $\mathcal{G}^{ij}_{2} (0)$ 
come from the diagrams illustrated in Fig.~\ref{fig:diagram-mueg}, which depend either linearly (left diagram) 
or quadratically (right diagram) on the lepton Yukawas, cf.~Eq.~\eqref{eq:yuk-alp}. 

Linear diagrams, which mimic the so-called Barr-Zee contributions~\cite{Barr:1990vd,Marciano:2016yhf}, are typically dominant
due to the logarithmic dependence on the ultraviolet cut-off.~\footnote{In ultraviolet complete models, the $c_{\gamma\gamma}$ arises only at loop-level, in such a way that this diagram would correspond to a two-loop contribution.} Their contribution to the dipole form factors reads
\begin{align}
\mathcal{F}_2^{ij}(0)_{\mathrm{lin.}}= - e^2 \dfrac{m_{\ell_j}^2}{8\pi^2\Lambda^2} \,a_{ij}^\ell\,c_{\gamma\gamma}\,g_{\gamma}(x_j)\,,
\label{eq:F2}\\
\mathcal{G}_2^{ij}(0)_{\mathrm{lin.}}= - e^2 \dfrac{m_{\ell_j}^2}{8\pi^2\Lambda^2} \,v_{ij}^\ell\,c_{\gamma\gamma}\,g_{\gamma}(x_j)\,,
\label{eq:G2}
\end{align}
where $x_j= m^2_a/m^2_{\ell_j} -i\eta$ (with $\eta \to 0^+$) and $g_\gamma(x)\approx 2 \log\Lambda^2/m^2_a$. The complete 
loop function is reported in Apendix \ref{app:loop_functions}.

By contrast, the quadratic contributions are finite, once self-energies are included, and depend on the flavor of the lepton $\ell_k$ running in the loop. The contributions involving only one LFV coupling 
(i.e.~with $k=i$ or $k=j$) read~\footnote{For generality, we have kept the contributions from a light lepton running in the loop in the 
second terms of Eq.~\eqref{eq:F2-0} and \eqref{eq:G2-0}. Nonetheless, these contributions turn out to be sub-dominant in most scenarios 
due to the suppression factor $m_{\ell_i}/m_{\ell_j}\ll 1$.} 
\begin{align}
\label{eq:F2-0}
\mathcal{F}^{ij}_2(0)_{\mathrm{quad.}} &= -\dfrac{m_{\ell_j}}{16\pi^2\Lambda^2}a_{ij}^\ell\Big{[}a_{jj}^\ell\,m_{\ell_j}\,g_1(x_j)+a_{ii}^\ell\,m_{\ell_i}\,g_2(x_j)\Big{]}\,,\\[0.3em]
\label{eq:G2-0}
\mathcal{G}^{ij}_2(0)_{\mathrm{quad.}} &= -\dfrac{m_{\ell_j}}{16\pi^2\Lambda^2}v_{ij}^\ell\Big{[}a_{jj}^\ell\,m_{\ell_j}\,g_1(x_j)-a_{ii}^\ell\,m_{\ell_i}\,g_2(x_j)\Big{]}\,.
\end{align}

\noindent In the $\mu \to e \gamma$ case, there is an additional contribution from the $\tau$-loop exchange which is induced by a double LFV source.
We find
\begin{align}
\label{eq:F2-1}
\mathcal{F}_{2}^{e\mu}(0)_{\mathrm{quad.}}  & = - \frac{m_{\mu}m_{\tau}}{32 \pi^{2} \Lambda^{2}}  \left( a_{e \tau}^\ell\, a_{\tau \mu}^\ell  - v_{e \tau}^\ell\, v_{\tau\mu}^\ell\right)\, g_{3}(x_\tau) \,,\\
\label{eq:G2-1}
\mathcal{G}_{2}^{e\mu}(0)_{\mathrm{quad.}} & = - \frac{m_{\mu}m_{\tau}}{32 \pi^{2} \Lambda^{2}} \left(v^\ell_{e \tau}\, a_{\tau\mu}^\ell -a_{e \tau}^\ell\, v_{\tau\mu}^\ell \right) \,   g_{ 3} (x_\tau) \,,
\end{align}

\noindent which show a $m_\tau/m_\mu$ enhancement compared to contributions involving a single LFV coupling. 
Similarly, $\tau \to \mu \gamma$ receives contributions from electron loops, which read
\begin{align}
\label{eq:F2-2}
\mathcal{F}_{2}^{\mu \tau}(0)_{\mathrm{quad}} &= -  \frac{ m_{\mu} m_{\tau} }{32 \pi^{2} \Lambda^{2}} \left(a^{\ell}_{\tau e } a^{\ell}_{\mu e} + v^{\ell}_{\tau e}  v^{\ell}_{\mu e}  \right) g_{4}(x_{\tau}) \,,\\
\label{eq:G2-2}
\mathcal{G}_{2}^{\mu \tau}(0)_{\mathrm{quad}} &= + \frac{ m_{\mu} m_{\tau} }{32 \pi^{2} \Lambda^{2}} \left( a^{\ell}_{\tau e }v^{\ell}_{\mu e} +  
v^{\ell}_{\tau e}  a^{\ell}_{\mu e} \right) g_{4}(x_{\tau}) \,,
\end{align}
while muon loops contribute to $\tau \to e \gamma$ as follows,
\begin{align}
\label{eq:F2-3}
\mathcal{F}_{2}^{e \tau}(0)_{\mathrm{quad}} &= -  \frac{ m_{\mu} m_{\tau} }{32 \pi^{2} \Lambda^{2}} 
\left(a^{\ell}_{\tau \mu } a^{\ell}_{e \mu} - v^{\ell}_{\tau \mu}  v^{\ell}_{e \mu}  \right) g_{4}(x_{\tau}) \,,\\
\label{eq:G2-3}
\mathcal{G}_{2}^{e \tau}(0)_{\mathrm{quad}} &=  -\frac{ m_{\mu} m_{\tau} }{32 \pi^{2} \Lambda^{2}} 
\left( a^{\ell}_{\tau \mu }v^{\ell}_{e \mu} -  v^{\ell}_{\tau \mu}  a^{\ell}_{e \mu} \right) g_{4}(x_{\tau}) \,,
\end{align}
with the loop functions $g_i(x)$ also collected in Appendix A.  We find full agreement with the results reported in Ref.~\cite{Bauer:2019gfk}.

In summary, the most general contributions to $\ell_j \to \ell_i \gamma$ with a single LFV coupling are given by the sum of the linear and quadratic contributions, 
see Eq.~\eqref{eq:F2}--\eqref{eq:G2-0}. 
Moreover, one should include the additional effects of Eq.~\eqref{eq:F2-1} and \eqref{eq:G2-1} in the $\mu \to e \gamma$ case, 
Eq.~\eqref{eq:F2-2} and \eqref{eq:G2-2} for $\tau \to \mu \gamma$, and finally Eq.~\eqref{eq:F2-3} and \eqref{eq:G2-3}
for $\tau \to e \gamma$.


\subsection{$\ell_{j} \to \ell_{i}\ell_k\ell_k$} 
\label{ssec:ndipole}

\begin{figure}[h!]
\centering
\includegraphics[width=0.65\textwidth]{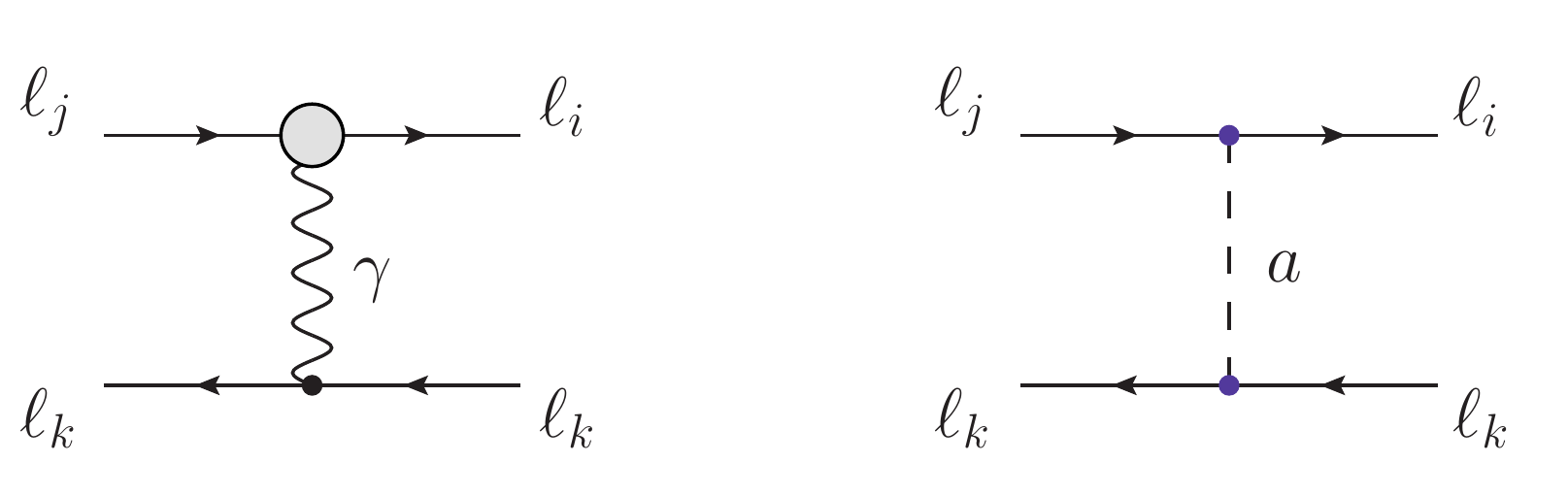}
\caption{\sl \small Diagrams illustrating the ALP contributions to {{$\ell_j\to \ell_{i}\ell_k\ell_k$}} at loop-level (left panel) and tree-level (right panel). The gray blob in the photonic contribution represents the different loop contributions illustrated in Fig.~\ref{fig:diagram-mueg}.}
\label{fig:diagram-mu3e}
\end{figure}

The processes $\ell_j \to \ell_{i}\ell_k\ell_k$ are described by the diagrams shown in Fig.~\eqref{fig:diagram-mu3e}. 
The ALP can contribute both at tree-level (right panel) or at one loop, via the effective $\ell_{j} \to \ell_{i} \gamma^{\ast}$ vertex (left panel).
Depending on the ALP mass, two different regimes arise: (i) for $m_a > m_{\ell_j}-m_{\ell_i}$ or $m_a< 2\, m_{{\ell_k}}$ the ALP is never produced on-shell, in such a way that there is a competition between tree and loop-level contributions, while (ii) for $2 \, m_{{\ell_k}} < m_a < m_{\ell_j}-m_{\ell_i}$ the ALP can be produced on-shell, making the tree-level exchange dominant. In the following we provide the relevant expressions in both cases and discuss the phenomenological implications. 

We start by parameterizing the general amplitude for the emission of an off-shell photon. In this case, Eq.~\eqref{eq:leff-dip} should be replaced by
\begin{equation}
\label{eq:leff-ndip}
i \mathcal{M}^{\mu}(\ell_j\to\ell_i \gamma^\ast)  =  i \, e \, \bar u_{i} (p-q) \Sigma^{\mu}_{ij}(q^2) u_{j}(p)\,,
\end{equation}
\noindent with
\begin{equation}
\label{eq:sigma-offshell}
\Sigma^{\mu}_{ij}(q^2) = \gamma_{\nu} \left(\mathcal F^{ij}_{1} (q^{2})  + \mathcal G^{ij}_{1} (q^{2}) \gamma^{5} \right)\left( g^{\mu \nu}  - \dfrac{q^{\mu} q^{\nu}}{q^2} \right) + \frac{  i \sigma^{\mu \nu}    q_{\nu} }{m_{\ell_j}} \left(\mathcal F^{ij}_{2} (q^{2}) + \mathcal G^{ij}_{2} (q^{2}) \gamma^{5}  \right)  \,,
\end{equation}
\noindent where the $\mathcal F^{ij}_{1,2} (q^{2})$ and $\mathcal G^{ij}_{1, 2} (q^{2})$ are form-factors depend on $q^2$ and on the masses of the particles running in the loops depicted in Fig.~\ref{fig:diagram-mueg}. 
The general expression for these functions, which are reported in Appendix~\ref{app:offshellff}, have been computed by independently using the packages {\sc{Feyncalc}}~\cite{Shtabovenko:2016sxi} and {\sc{Package-X}}~\cite{Patel:2016fam}. We verified that these expressions coincide 
with the results given in Sec.~\ref{ssec:dipole} in the limit $q^2\to 0$. In particular, $\mathcal{F}_1^{ij}(0)=\mathcal{G}_1^{ij}(0)=0$, as expected by gauge invariance. 

Even though the form-factors reported in Appendix~\ref{app:offshellff} provide the most general description of the transition $\ell_j\to \ell_i \gamma^\ast$, 
it is convenient to derive simplified expressions which are valid for off-shell ALPs, i.e. for $m_a > m_{\ell_j}-m_{\ell_i}$, and which are more convenient for phenomenological analyses. In this case, $\mathcal{F}_{1}^{ij}$ 
and $\mathcal{G}_{1}^{ij}$ are well-approximated by a series around $q^2=0$,
\begin{align}
\label{eq:dF10}
\begin{split}
\mathcal{F}_1^{ij} (q^2)&= q^2\,\dot{\mathcal{F}}_1^{ij}(0)+ \mathcal{O}(q^4)\,,\\[0.3em]
\mathcal{G}_1^{ij} (q^2)&= q^2\,\dot{\mathcal{G}}_1^{ij}(0)+ \mathcal{O}(q^4)\,,
\end{split}
\end{align}
where $\dot{\Phi} \equiv \mathrm{d}\Phi/\mathrm{d}q^2$. Similarly, the dipole form-factors $\mathcal{F}_2^{ij}(q^2)$ and $\mathcal{G}_2^{ij}(q^2)$ are well 
described at leading order by setting $q^2=0$. The complete expressions for $\mathcal{F}_{2}^{ij}(0)$, $\dot{\mathcal{F}}_{1}^{ij}(0)$, $\mathcal{G}_{2}^{ij}(0)$ and $\dot{\mathcal{G}}_{1}^{ij}(0)$ are 
reported in Appendix \ref{app:taylor}.

The one-loop contributions computed above can now be combined with the tree-level one (see~Fig.~\ref{fig:diagram-mu3e}) to provide the general expression for $\Gamma(\ell_j \to \ell_{i}\ell_k\ell_k)$. For compactness, form factors are expanded around $q^2=0$, as described above. 

\paragraph*{Off-shell decay rates} We first consider the off-shell scenario with $m_a > m_{\ell_j}-m_{\ell_i}$. The 
{$\ell_{j} \to \ell_{i}\ell_k\ell_k$} decay rate in this regime can be decomposed in three pieces, namely (i) the photonic contribution, 
(ii) the ALP-mediated tree-level exchange and (iii) their interference, namely
\begin{align}
\Gamma(\ell_j \!\to\! \ell_i \ell_k \ell_k) &=  \Gamma(\ell_j \!\to\! \ell_i \gamma^\ast \!\to\! \ell_i \ell_k \ell_k) + 
\Gamma(\ell_j \!\to\! \ell_i a^\ast \!\to\! \ell_i \ell_k \ell_k)+
\delta_{ik} \, \Gamma(\ell_j \!\to\! 3 \ell_i)_{\mathrm{int.}}\,.
\end{align}
\noindent We compute each of these contributions by keeping the leading $q^2$-dependence in the one-loop form factors. For the photonic contribution, we find  
\begin{align}
\label{eq:Gamma-loop}
\Gamma(\ell_j&\!\to\! \ell_i \gamma^\ast  \!\to\! \ell_i \ell_k \ell_k) = 
\dfrac{\alpha_{\mathrm{em}}^2 m_{\ell_j}}{6\pi} 
\Bigg{\lbrace} \! \left(\! \log \dfrac{m_{\ell_j}^2}{m_{\ell_i}^2} - 3 + \dfrac{\delta_{ik}}{4} \! \right)\Big{[} 
|F_2^{ij}(0)|^2+|G_2^{ij}(0)|^2\Big{]}  \\
&+ \left(\! 1 + \frac{\delta_{ik}}{2} \! \right)\!\left[ \dfrac{m_{\ell_j}^4}{4} \left(|\dot{F}_1^{ij}(0)|^2+|\dot{G}_1^{ij}(0)|^2\right) - 
m_{\ell_j}^2 \mathrm{Re}\Big{(} \dot{F}_1^{ij}(0) F_2^{ij}(0)^\ast{-\dot{G}_1^{ij}(0) G_2^{ij}(0)^\ast}\Big{)}
\right]\!\Bigg{\rbrace}\,,
\nonumber
\end{align}
where we have used $m_{\ell_j}\gg m_{\ell_i}$, finding agreement with the standard expressions available in the literature~\cite{Kuno:1999jp}. 
For the tree-level term, we obtain the following expression by neglecting the ALP width,
\begin{equation}
\label{eq:Gamma-tree}
\Gamma(\ell_j\to \ell_i a^\ast \to \ell_i \ell_k \ell_k) =  \dfrac{|a^\ell_{kk}|^2\, |s^\ell_{ij}|^2}{32 \pi^3}\, \dfrac{m_{\ell_j}^3 m_{\ell_k}^2}{\Lambda^4}\, 
\varphi^{ik}_{0} (x_j)\,,
\end{equation}
where $x_j=m_a^2/m_{\ell_j}^2$, as before, and the phase-space function {{$\varphi^{ik}_{0}(x)$}} is reported in Appendix \ref{app:mu3e-kin}. 
In the limit of large ALP masses, this function satisfies {{$\varphi^{ik}_{0}(x) \propto 1/x^2+\mathcal{O}(1/x^3)$}}, in agreement with the decoupling limit. Similarly, we computed the interference of both contributions, which is given by
\begin{align}
\label{eq:Gamma-int}
\begin{split}\Gamma(\ell_j\to 3\,\ell_i)_\mathrm{int.} = \dfrac{\alpha_{\mathrm{em}} }{96 \pi^2}\dfrac{m_{\ell_j}^2 m_{\ell_i}}{\Lambda^2} 
\, &\Bigg{\lbrace}-2\,\varphi_1(x_j)\,\mathrm{Re}\Big{[}a_{11}^\ell \,\Big{(}a_{21}^\ell \, {F}_2(0)^\ast - v_{21}^\ell \, {G}_2(0)^\ast\Big{)}\Big{]} \\&\;\;+m_{\ell_j}^2\,\varphi_2(x_j)\,\mathrm{Re}\Big{[}a_{11}^\ell \,\Big{(}a_{21}^\ell \, \dot{F}_1(0)^\ast + v_{21}^\ell \, \dot{G}_1(0)^\ast\Big{)} \Big{]} \Bigg{\rbrace}\,,
\end{split}
\end{align}
\noindent where the phase-space functions satisfy $\varphi_{1,2}(x) = 1/x+\mathcal{O}(1/x^2)$, with the complete functions collected 
in Appendix \ref{app:mu3e-kin}. Note that the interference between tree and loop-level contributions vanishes identically for $k\neq i$.
The phenomenological implication of these expressions are discussed in Sec.~\ref{ssec:leptonic-numerics}.
 
In the off-shell scenario with $m_a<2\,m_{\ell_{k}}$,  Eq.~\eqref{eq:Gamma-loop} remains the same, while Eq.~\eqref{eq:Gamma-tree} and \eqref{eq:Gamma-int} should be reevaluated with the appropriate phase-space integration. In this case, it is more difficult to provide a compact analytical expression as there are more mass scales involved. In the phenomenological analysis, we integrate the form factors reported in Appendix~\ref{app:offshellff} numerically. We have also checked that the form factor expansion around $q^2=0$ remains a reasonable approximation.

\paragraph*{On-shell decay rates} The above expressions can be simplified in the case where the ALP is produced on-shell, i.e.~for  
$2m_{{\ell_k}} < m_a < m_{\ell_j}-m_{\ell_i}$. In this case, the interference term in Eq.~\eqref{eq:Gamma-int} becomes negligible, while the 
photonic contribution remains identical to Eq.~\eqref{eq:Gamma-loop}. On the other hand, the tree-level ALP exchange can be described in the narrow-width approximation, 

\begin{equation}
\label{eq:mu3e-onshell}
\Gamma(\ell_j\to \ell_i a \to \ell_i \ell_k \ell_k) \approx \Gamma(\ell_j\to \ell_i a)\,\mathcal{B}(a\to \ell_k\ell_k)\,,
\end{equation}
\noindent with
\begin{align}
\label{eq:muea}
\Gamma(\ell_j\to \ell_i a) = \dfrac{m_{\ell_j}^3}{16\pi}\left(1-\dfrac{m_a^2}{m_{\ell_j}^2}\right)^2 \dfrac{|s_{ij}^\ell|^2}{\Lambda^2}\,,
\end{align}
where we have used that $m_{\ell_j}\gg m_{\ell_i}$, and
\begin{align}
\label{eq:alilj}
\mathcal{B}(a\to\ell_k^{-} \ell_k^{+}) &= \tau_a\,\dfrac{|a_{kk}^\ell|^2}{\Lambda^2}\dfrac{m_a m_{\ell_k}^2}{2\pi}\sqrt{1-\dfrac{4 m_{\ell_k}^2}{m_a^2}}\,,
\end{align}

\noindent where $\tau_a$ denotes the ALP lifetime.

To assess the limits on ALP couplings in the on-shell regime, one should estimate the ALP flight distance and verify that it decays inside the detector, as shall be discussed next.
The ALP boosted decay length in the lab frame is given by
\begin{equation}
l_a = \dfrac{c\,|p_a|}{m_a\, \Gamma_a}\,,
\label{eq:alp_decay_lenght}
\end{equation}
where $p_a$ denotes the ALP momentum in the lab frame. By assuming that $\ell_j$ decays at rest, as in the case of $\mu\to e$ experiments 
such as MEG~\cite{TheMEG:2016wtm} or Mu3e~\cite{Bartoszek:2014mya}, $|p_a|$ reads
\begin{equation}
\label{eq:pa-rest-frame}
|p_a| = \dfrac{\lambda^{1/2}(m_a,m_{\ell_j},m_{\ell_i})}{2\, m_{\ell_j}}\,,
\end{equation}
with $\lambda(a,b,c)\equiv(a^2-(b-c)^2)(a^2-(b+c)^2)$. In our numerical analysis, we will naively impose the relaxed bound that $\ell_a$ is not larger than $\approx 1$~m in order for the ALP decay to be considered prompt. A more refined analysis can be performed in experimental searches, by using the displaced vertex as a tool to set even more stringent limits than searches based on prompt decays~\cite{Heeck:2017xmg}.~\footnote{
Similar displaced-vertex searches have been recently performed by LHCb in the decays $B\to K^{(\ast)} a \to K^{(\ast)} \mu\mu$~\cite{Aaij:2015tna,Aaij:2016qsm}, which provide some of the most stringent limits on GeV ALPs with couplings to both quarks and leptons~\cite{Gavela:2019wzg}.} Another limitation of our analysis is the reinterpretation of $\tau\to e$ and $\tau\to \mu$ limits.
Indeed, since $\tau$'s are not produced at rest in current experiments, Eq.~\eqref{eq:pa-rest-frame} does not apply in this case. 
The correct assessment of $\tau$ LFV limits in the resonant region would require a dedicated experimental study, as already suggested in Ref.~\cite{Heeck:2017xmg}

\subsection{$\ell_{j} \to \ell_i\,\gamma\gamma$ }

\begin{figure}[h!]
\centering
\includegraphics[width=0.29\textwidth]{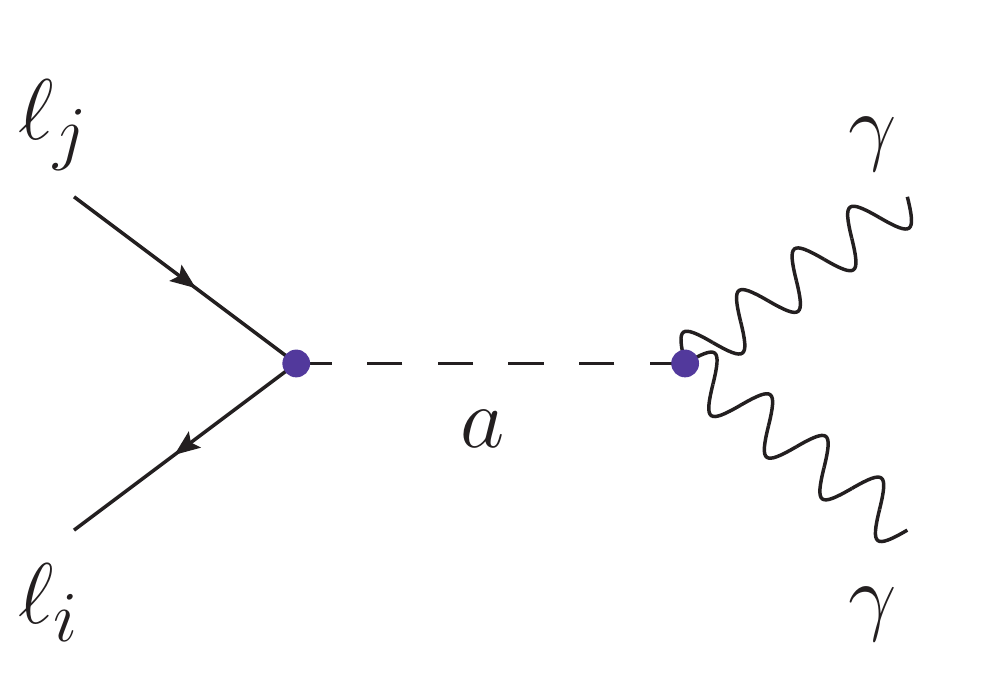}
\caption{\sl \small Dominant contribution to $\ell_j\to \ell_i \gamma\gamma$ for light ALPs, i.e.~$m_a \lesssim m_{\ell_j}$.}
\label{fig:diagram-muegg}
\end{figure}

The next purely leptonic decay mode we discuss is $\ell_j \to \ell_i \gamma\gamma$. We focus on the tree-level contribution illustrated in Fig.~\ref{fig:diagram-muegg}, which is the dominant one for light ALPs. Once again, we separate the off-shell region, $m_a > m_{\ell_j}-m_{\ell_i}$, from 
the on-shell one, $m_a < m_{\ell_j}-m_{\ell_i}$.

\paragraph*{Off-shell decay rates} 
The general expression for the branching ratio of $\ell_j \to \ell_i \gamma\gamma$ in the off-shell regime ($m_a > m_{\ell_j}$) is given by
\begin{equation}
\mathcal{B}(\ell_j \to \ell_i \gamma\gamma) =  \tau_{\ell_i}\,\dfrac{e^4\,c_{\gamma\gamma}^2|s_{ij}^\ell|^2}{\Lambda^4} \dfrac{m_{\ell_i}^5}{192 \pi^3}\, \varphi (x)\,,
\end{equation}
where $x=m_a^2/m_{\ell_j}^2$ and the loop-function $\varphi(x)$ reads
\begin{equation}
\varphi(x) = 1 + 12 x (x -1) + 6 x\left(2 x^2-3 x+ 1\right) \log \frac{x-1}{x}\,,
\end{equation}
with $\varphi(x) = 1/(10\,x^{2}) + \mathcal{O}(1/x^{3})$ as $x\gg 1$.
%

\paragraph*{On-shell decay rates} 
In the on-shell regime, the branching ratio of $\ell_{j} \to \ell_i\,\gamma\gamma$ can be obtained exploiting the narrow-width approximation, 
leading to the following result
\begin{align}
\begin{split}
\mathcal{B}(\ell_j \to \ell_i \gamma\gamma) &\approx \mathcal{B}(\ell_j \to \ell_i a) \, \mathcal{B}(a\to\gamma\gamma)\,,
\end{split}
\end{align}
\noindent where
\begin{align}
\label{eq:agammagamma}
\mathcal{B}(a\to\gamma\gamma) &= \tau_a\, \dfrac{e^4|c_{\gamma\gamma}|^2}{\Lambda^2}\dfrac{m_a^3}{4\pi}\,,
\end{align}
\noindent and the leptonic decay rate is given in Eq.~\eqref{eq:muea}. As discussed in Sec.~\ref{ssec:leptonic-numerics}, given the present experimental constraints, the process $\ell_{j} \to \ell_i\,\gamma\gamma$ turns out to be effective 
in limiting the ALP parameter space only in the on-shell regime. \\

\subsection{$\mu^-\to e^-$ conversion in nuclei}
\label{ssec:nuclei}

We now discuss $\mu^-\to e^-$ conversion in nuclei, which will become one of the most sensitive LFV probes in the coming years (cf.~Ref.~\cite{Calibbi:2017uvl} 
for an overview of experimental prospects). We consider this observable to be a purely leptonic probe, since ALP couplings to quarks and gluons do not induce coherent contributions to the conversion rate, being therefore negligible~\cite{Kuno:1999jp}. The only relevant contributions are then the ones stemming from the (off-shell) photon penguins computed in Sec.~\ref{ssec:ndipole}.~\footnote{See Ref.~\cite{Cirigliano:2017azj} for an estimation of spin-dependent contributions to these processes arising from axial-vector and tensor operators.} 

The conversion branching fraction is defined as the ratio of the $\mu\to e$ conversion rate over the nuclear capture one. Following Ref.~\cite{Kuno:1999jp}, 
we have
\begin{equation}
\mathcal{B}(\mu^{-}N\rightarrow e^{-}N) = \frac{ 8\alpha^{5}m_{\mu}
Z^{4}_{\mathrm{eff}} Z F_{p}^2 }{\Gamma_{\mathrm{capt}}} \,  \xi^2\,,
\label{eq:wein}
\end{equation}
where $Z_\mathrm{eff}$ is the effective atomic charge, $F_p$ parameterizes the nuclear matrix element and $\Gamma_{\mathrm{capt}}$ stands for the total muon capture rate. The factor $\xi^{2}$ accounts for the photon-exchange contributions as follows
\begin{align}
\begin{aligned}
\xi^2 &=  \left| \mathcal{F}_{1}^{\mu e}(-m_{\mu}^2) + \mathcal{F}_{2}^{\mu e}(-m_{\mu}^2) \right|^2 
+ \left|\mathcal{G}_{1}^{\mu e}(-m_{\mu}^2) - \mathcal{G}_{2}^{\mu e }(-m_{\mu}^2) \right|^2  \,,
\label{eq:form_mue}
\end{aligned}
\end{align}
where $\mathcal{F}_{1,2} (-m_{\mu}^{2})$ and $\mathcal{G}_{1,2} (-m_{\mu}^{2})$ are the $\mu \to e \gamma^{\ast}$ form factors evaluated at $q^{2} = -m_{\mu}^{2}$, cf.~Eq.~\eqref{eq:sigma-offshell} and Appendix~\ref{app:offshellff}. For $m_{a}>m_{\mu}$, evaluating the form-factors at $q^2=0$ is a very good approximation. In particular,  considering only one flavor-violating coupling and neglecting 
the electron mass, this expression simplifies to
\begin{align}
\xi^{2} \propto |v_{\mu e}^\ell|^{2} + |a_{\mu e}^\ell|^{2} = |s_{\mu e}^\ell|^2\,,
\end{align}
as in the LFV quantities discussed above. In the specific case of gold ($^{197}_{79} \mathrm{Au}$) and aluminum ($^{27}_{13} \mathrm{Al}$) atoms, the necessary inputs are given respectively by \cite{Kitano:2002mt},
\begin{align}
&Z = 79\,, \qquad Z_{\mathrm{eff}}= 33.5\,, \qquad F_{p} = 0.16\,,\qquad \Gamma_{\mathrm{capt}} = 8.59868 \cdot 10^{-18} \, \mathrm{GeV}\,,\\[0.4em]
&Z = 13\,, \qquad Z_{\mathrm{eff}}= 11.5\,, \qquad F_{p} = 0.64\,,\qquad \Gamma_{\mathrm{capt}} = 4.64079\times 10^{-19} \, \mathrm{GeV}\,,
\end{align}
which can be replaced in Eq.~\eqref{eq:wein} to give 
\begin{align}
\mathcal{B}^{\mathrm{th}}(\mu^{-}{\mathrm{Au}\rightarrow e^{-}{\mathrm{Au}}}) &\approx 5.2 \times 10^{12} \,  \xi^{2}\,,\qquad\quad
\mathcal{B}^{\mathrm{th}}(\mu^{-}\mathrm{Al}\rightarrow e^{-}\mathrm{Al}) \approx 3.52 \times 10^{12} \,  \xi^{2}\,.
\end{align}
Currently, the most stringent limit is $\mathcal{B}(\mu^{-}{\mathrm{Au}\rightarrow e^{-}{\mathrm{Au}}})<7\times 10^{-13}$, obtained by the Sindrum-II collaboration~\cite{Bertl:2006up}. In the near future, the Mu2e experiment at Fermilab~\cite{Bartoszek:2014mya} and Comet~\cite{Adamov:2018vin} at J-PARC aim to improve the experimental sensitivity to $\mathcal{O}(10^{-17})$ by using aluminum atoms, cf.~Table~\ref{tab:exp-lep}.

\subsection{Numerical results and discussion}
\label{ssec:leptonic-numerics}

In this section we discuss the phenomenological implications of the results obtained above. Our main goal is to explore the complementarity 
of the different leptonic probes and their experimental prospects at current/future experimental facilities.

To derive constraints from existing data, we focus on a benchmark scenario defined by lepton-flavor universal couplings to leptons 
($a^\ell \equiv a_{ii}^\ell$), with the other flavor-conserving couplings vanishing at tree-level. In this case, the effective ${a\gamma\gamma}$ 
coupling is generated by the (irreducible) one-loop contributions from leptonic couplings~\cite{Bauer:2017ris},

\begin{equation}
\label{eq:cagammagamma}
c_{\gamma\gamma}^{\text{loop}} = \dfrac{a^\ell}{8\pi^2}\,\sum_{i=e,\mu,\tau} B_{1}(\tau_i)\,,
\end{equation}
where $\tau_i= 4 m_i^2/m_a^2-i\,\eta$ and the loop-function is given by
\begin{equation}
B_1(\tau)=1 - \tau \,f^2(\tau)\qquad \mathrm{with} \qquad f(\tau)=\left\{\begin{matrix}
\mathrm{arcsin}\frac{1}{\sqrt{\tau}},\hspace*{4.em} \tau \geq 1\\ 
\frac{\pi}{2}+\frac{i}{2}\log\frac{1+\sqrt{1-\tau}}{1-\sqrt{1-\tau}} ,\hspace*{1em} \tau<1
\end{matrix}\right.\,,
\end{equation}
\noindent which behaves as $B_1(\tau) \approx 1/\sqrt{\tau}$ in the $\tau \to \infty$ asymptotic limit, in such a way that only fermions lighter than $m_a$ contribute significantly to Eq.~\eqref{eq:cagammagamma}. To derive the relevant constraints to this scenario, it is crucial to determine the ALP total width, $\Gamma_a$, and its flight distance in a given experiment~\cite{Heeck:2017xmg}, as discussed above. The constraints on $|a_{\mu e}^\ell|$, $|a_{\tau e}^\ell|$ and $|a_{\tau \mu}^\ell|$ are shown in Fig.~\ref{fig:mue-constraints}, as a function of the ALP mass $m_a$, by setting the flavor-conserving coupling to $a^\ell=1$. Several comments are in order:
\begin{itemize}
\item Out of the whole set of $\mu\to e$ processes, $\mathcal{B}(\mu\to 3e)$ and $\mathcal{B}(\mu\to e\gamma\gamma)$ impose the most stringent bounds for $20~\mathrm{MeV}<m_a \lesssim 100~\mathrm{MeV}$ due to the resonant-enhancement of the branching fractions. The upper bound corresponds to the kinematical threshold of on-shell ALP production, i.e.~$m_a < m_\mu-m_e$, 
while the lower bound $m_a\gtrsim 20~$MeV comes from the requirement that the ALP decays inside the detector, i.e.~$\ell_a \lesssim 1~$m, cf.~Eq.~\eqref{eq:alp_decay_lenght}.
\item For $m_a \lesssim 20~$MeV, the ALP decays outside of the detector, in such a way that $\mu\to e a$ is indistinguishable from 
$\mu\to e +\mathrm{inv.}$, from which we obtain the most stringent constraint in this mass range. In principle, these limits could be improved if dedicated experimental searches for displaced vertices are performed~\cite{Heeck:2017xmg}. 
\item For $m_a > m_\mu$, the process $\mathcal{B}(\mu\to 3e)$ is dominated by the loop exchange of ALPs. Indeed, while tree-level effects decouple with inverse powers of $m_a$, loop-induced contributions exhibit a smoother logarithmic dependence. 
For this reason, $\mu\to e\gamma$ is currently the most constraining process for these masses. Currently, $\mu \to e$ conversion in nuclei 
is less sensitive to ALP couplings, since $\mathcal{B}(\mu N\to e N) \approx \mathcal{B}(\mu \to 3e)  \approx 10^{-2} \times \mathcal{B}(\mu\to e\gamma)$. This is expected to change in the future thanks to the Mu2E \cite{Bartoszek:2014mya} and COMET \cite{Adamov:2018vin} experiments, which will improve the present sensitivity by orders of magnitude, cf.~Table~\ref{tab:exp-lep}.

\begin{table}[!p]
\centering
  \renewcommand{\arraystretch}{1.5} 
\begin{tabular}{|c|ccc|}
\hline
Decay mode & Exp.~limit & Future prospects & Ref.\\
\hline\hline
$\mu\to e\gamma$ & $4.2\times 10^{-13}$ 	& $\approx 6\times 10^{-14}$ & \cite{Tanabashi:2018oca} 	\\
$\mu\to 3e$ & $1.0\times 10^{-12}$	& $\approx 10^{-16}$ & \cite{Tanabashi:2018oca} 	\\
$\mu\to e\gamma\gamma$ & $7.2\times 10^{-11}$	& -- & \cite{Tanabashi:2018oca} 	\\
$\mu\to e+\mathrm{inv}$ & $\approx 10^{-5}$	& -- & \cite{Bayes:2014lxz} 	\\
$\mu^-\,\mathrm{Ti}\to e^-\,\mathrm{Ti}$ & $4.3\times 10^{-12}$	& -- & \cite{Dohmen:1993mp} \\
$\mu^-\,\mathrm{Au}\to e^-\,\mathrm{Au}$ & $7\times 10^{-13}$	& $\approx 10^{-17}$ & \cite{Bertl:2006up} \\
$\mu^-\,\mathrm{Al}\to e^-\,\mathrm{Al}$ & --	& $\approx 10^{-17}$ & \cite{Bartoszek:2014mya,Adamov:2018vin} \\
\hline
\hline 
$\tau\to e \gamma$	& $3.3\times 10^{-8}$	& $\approx 3\times 10^{-9}$ & \cite{Tanabashi:2018oca} 	\\
$\tau\to 3e$	&  $2.7\times 10^{-8}$	& $\approx 5 \times 10^{-10}$ & \cite{Tanabashi:2018oca} 	\\
$\tau \to e \mu^+\mu^-$ & $1.7\times 10^{-8}$ & $\approx 6 \times 10^{-10}$ & \cite{Tanabashi:2018oca}  \\
$\tau\to e +\mathrm{inv}$	& 	$\approx 5\times 10^{-3}$ & -- & \cite{Albrecht:1995ht} 	\\
\hline
\hline
$\tau\to \mu \gamma$	&  $4.4\times 10^{-8}$	& $\approx 10^{-9}$ & \cite{Tanabashi:2018oca} 	\\
$\tau\to 3\mu$	& $2.1\times 10^{-8}$	&  $\approx 4 \times 10^{-10}$& \cite{Tanabashi:2018oca} 	\\
$\tau\to \mu e^+e^-$	& $1.8 \times 10^{-8}$	& $\approx 3 \times 10^{-10}$ & \cite{Tanabashi:2018oca} 	\\
$\tau\to \mu +\mathrm{inv}$	& $\approx 5\times 10^{-3}$	& -- & \cite{Albrecht:1995ht} 	\\
\hline
\end{tabular}
\caption{\em \small Most relevant experimental limits on purely leptonic LFV processes and future prospects for MEG-II~\cite{Baldini:2013ke}, Mu2E~\cite{Bartoszek:2014mya}, Mu3E~\cite{Blondel:2013ia}, COMET~\cite{Adamov:2018vin} and Belle-II \cite{Kou:2018nap}.}  
\label{tab:exp-lep}
\end{table}

\begin{figure}[p!]
\centering
\includegraphics[width=0.5\textwidth]{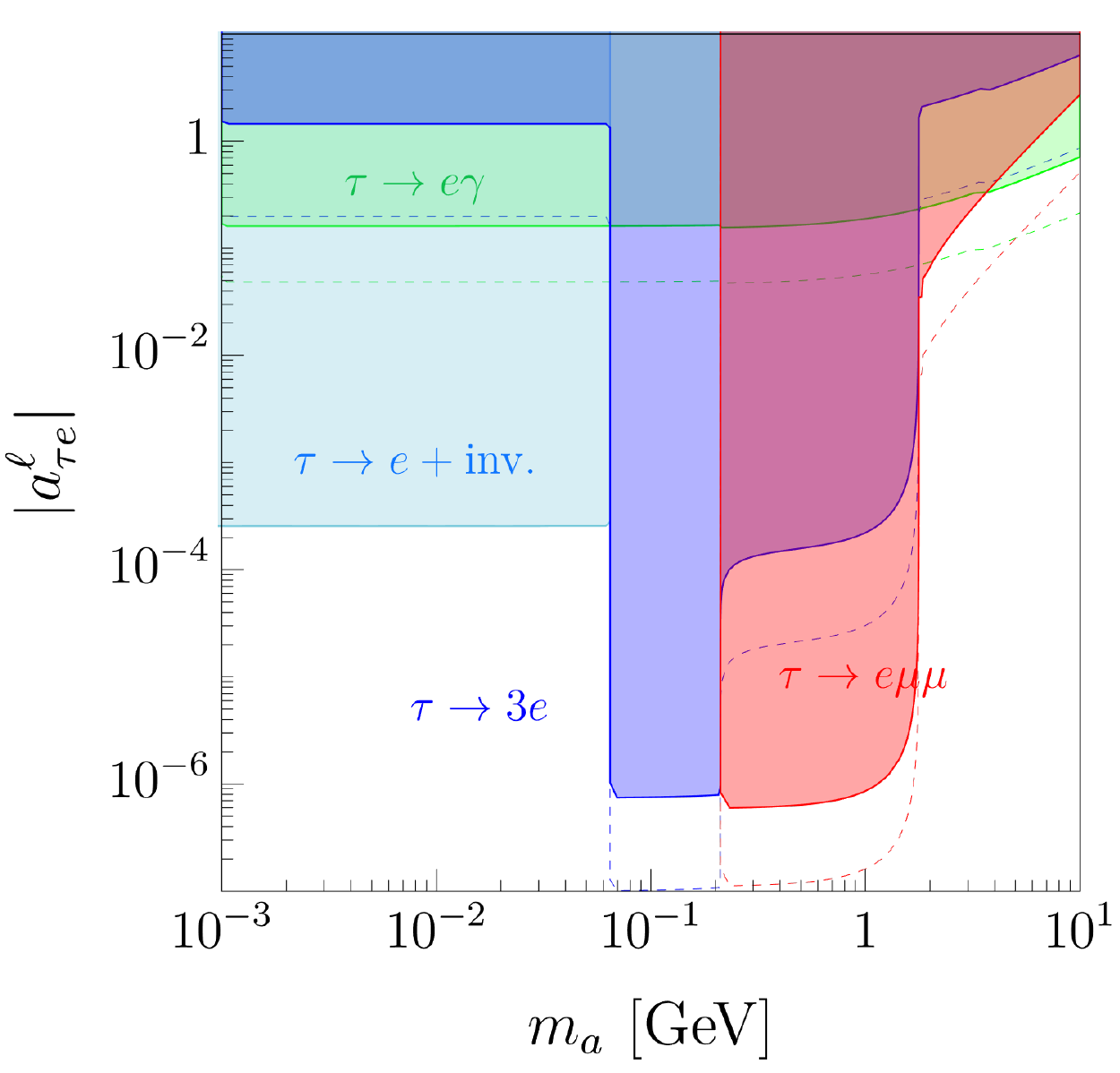}~\includegraphics[width=0.5\textwidth]{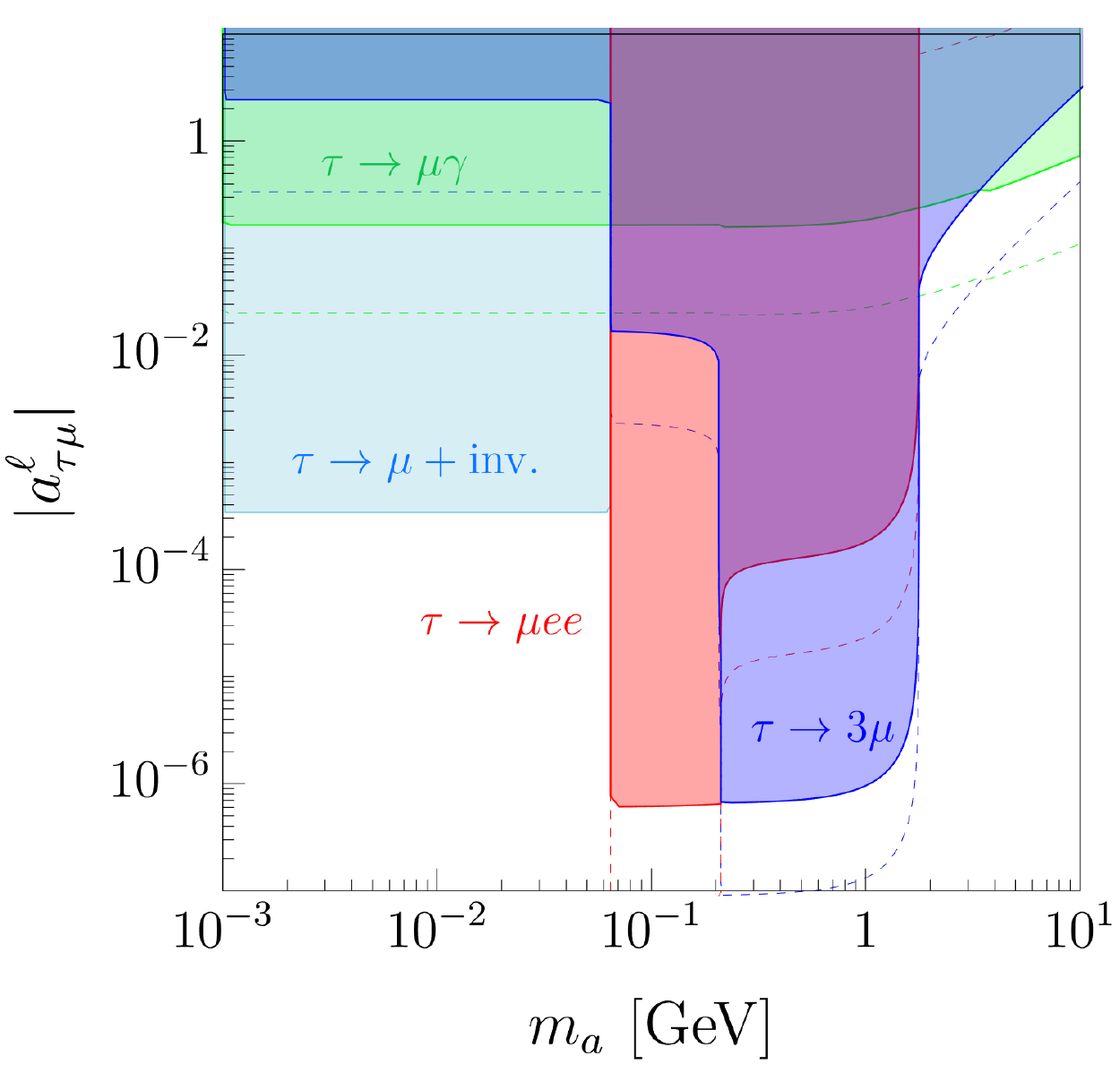}\\[0.9em]
\includegraphics[width=0.5\textwidth]{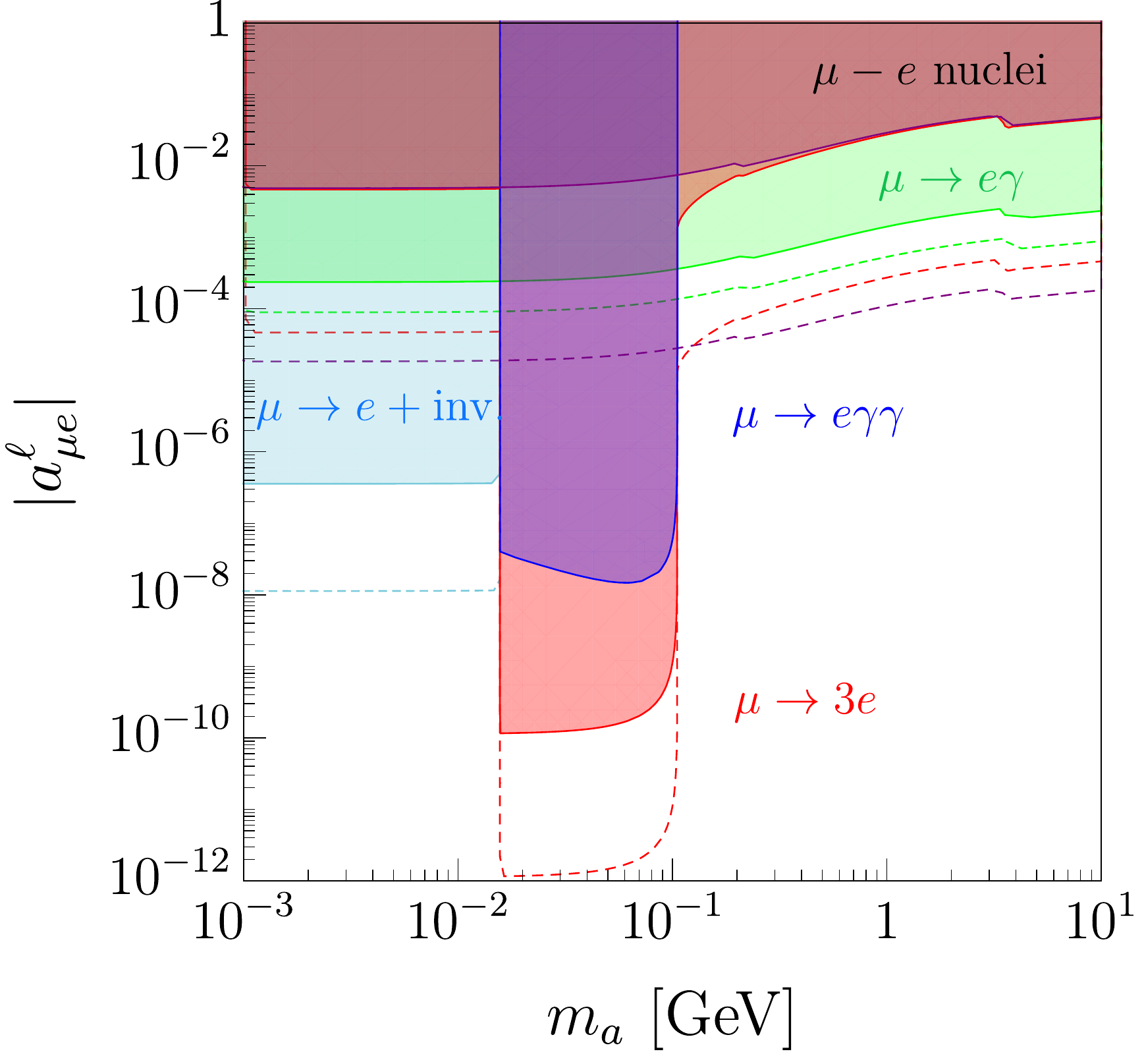}
\caption{\small\sl Constraints on $|s_{\mu e}^\ell|$ (lower panel), $|s_{\tau e}^\ell|$ (upper left panel) and $|s_{\tau \mu}^\ell|$ (upper right panel) as a function of the ALP mass 
derived from the experimental bounds listed in Table~\ref{tab:exp-lep}. Dashed lines correspond to future experimental prospects. We consider a benchmark scenario where $a_{ii}^\ell=1$ 
are the only flavor-conserving tree-level couplings, while $c_{\gamma\gamma}$ is induced at one-loop level, cf.~Eq.~\eqref{eq:cagammagamma}.  The other couplings are neglected in our analysis.}
\label{fig:mue-constraints}
\end{figure}

\item Bounds on $\tau\to\mu$ and $\tau\to e$ transitions share similar features to the ones already outlined above for the $\mu\to e$ transition. An important difference stems from the interplay between constraints from ALP on-shell production and the requirement that the ALP decays inside the detector. In particular, the three body decays $\tau\to 3\mu$ and $\tau\to e\mu\mu$ set the most stringent bounds in the ranges $ 2\,m_\mu < m_a < m_\tau-m_\mu$ and $ 2\, m_\mu < m_a < m_\tau-m_e$, respectively, which correspond precisely to the kinematical thresholds for on-shell ALP production. On the other hand, $\tau\to \mu ee$ and $\tau\to 3e$ are the most sensitive processes in the region $7~\mathrm{MeV}\lesssim m_a < 2m_\mu$, where the lower limit stems from the ALP lifetime constraint. 
Note, in particular, that our flavor ansatz is such that the relation $\mathcal{B}(\tau \to 3\,\mu)/\mathcal{B}(\tau \to \mu ee)=\mathcal{B}(\tau \to e\mu\mu)/\mathcal{B}(\tau \to 3e)=m_\mu^2/m_e^2$ holds exactly in the region where these observables are resonantly enhanced. 
\end{itemize}
%

\begin{figure}[t]
\centering
\includegraphics[width=0.5\textwidth]{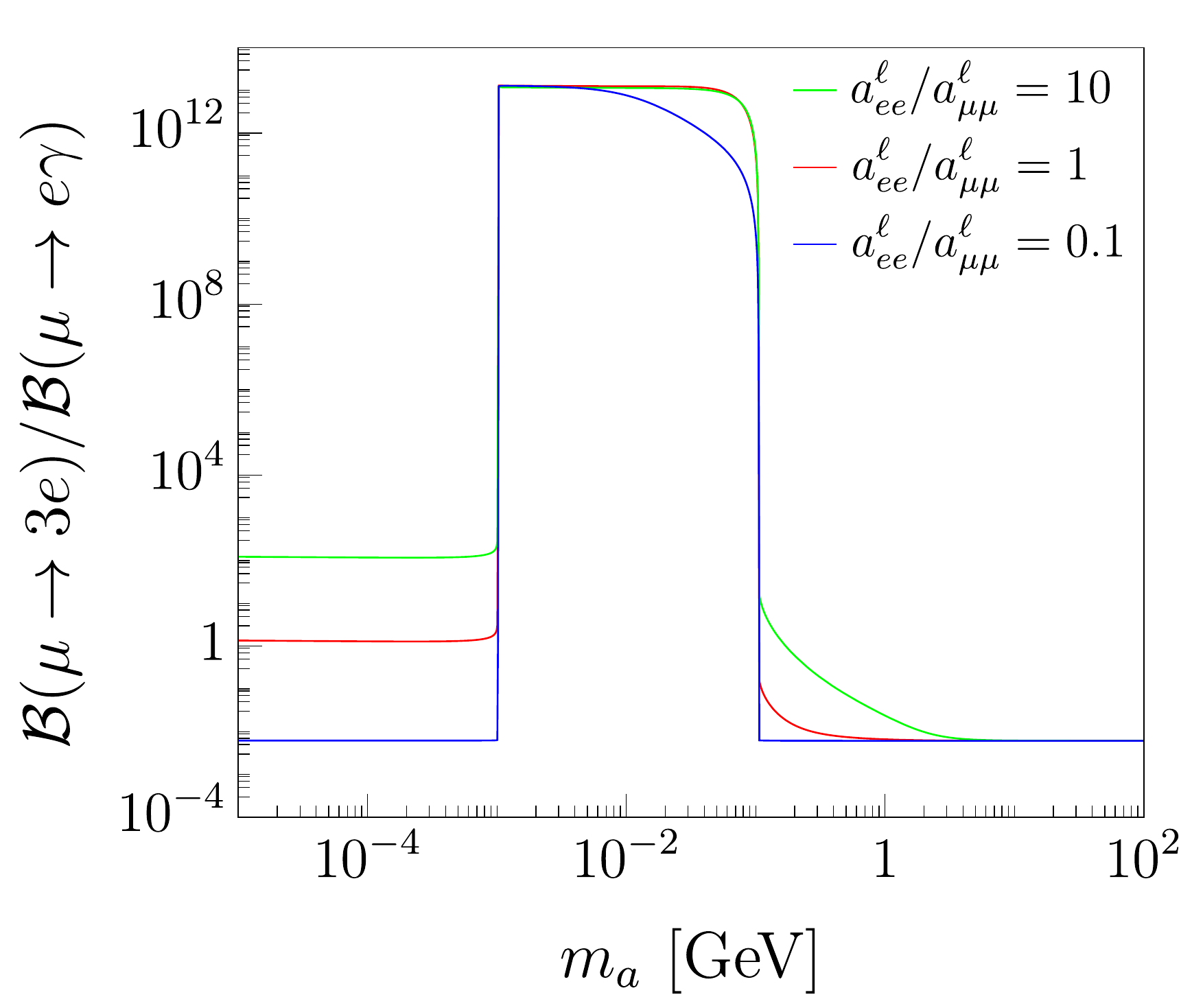}~\includegraphics[width=0.5\textwidth]{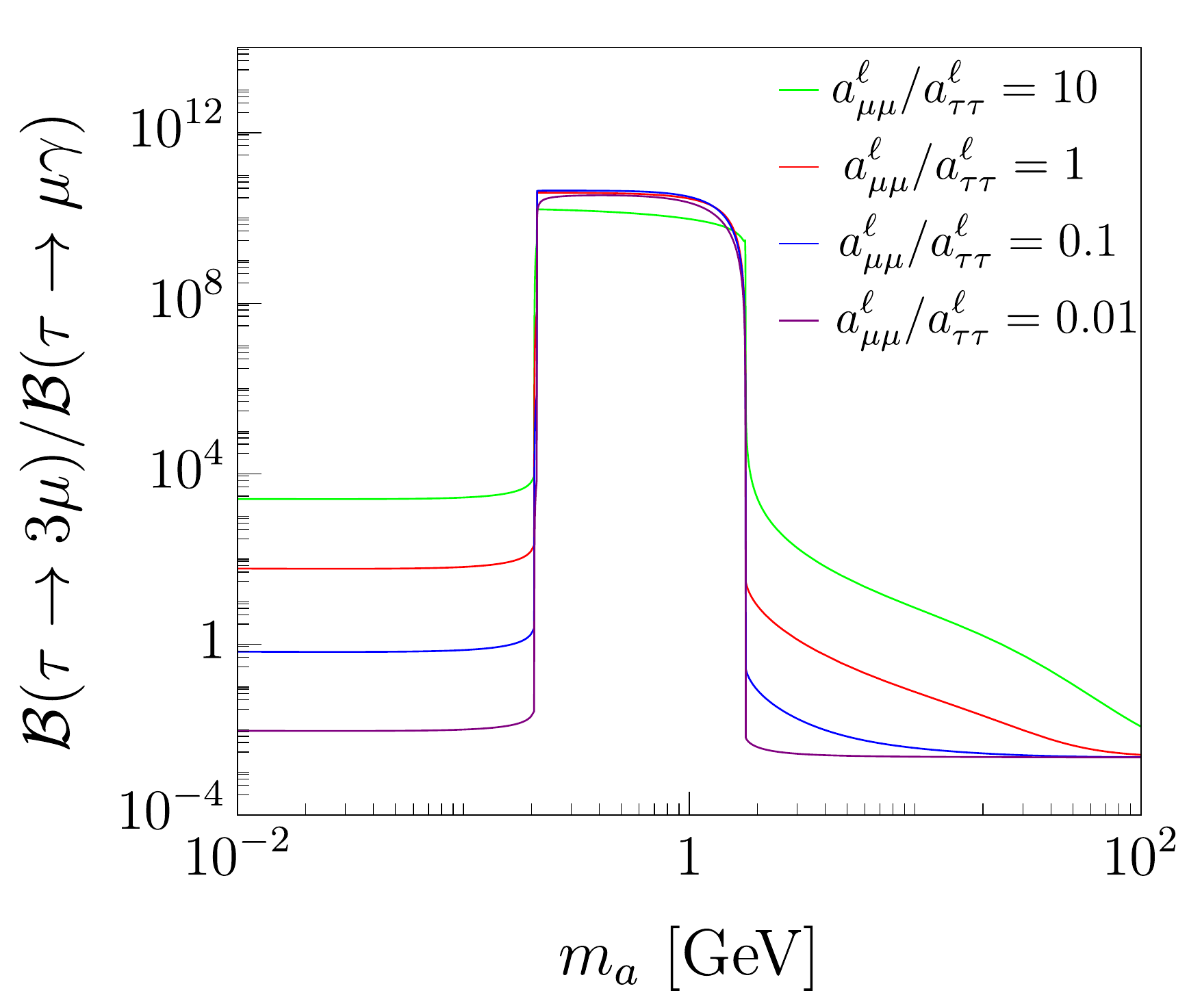}
\caption{\small\sl $\mathcal{B}(\mu \to 3 e)/\mathcal{B}(\mu\to e\gamma)$ and $\mathcal{B}(\tau \to 3\mu)/\mathcal{B}(\tau\to\mu\gamma)$ 
as a function of $m_a$ for the benchmark ratios $a^\ell \equiv a_{ii}^\ell$ specified in the plot. The coupling $c_{\gamma\gamma}$ is 
assumed to be induced at one-loop. See the text for more details.}
\label{fig:mue-illustration}
\end{figure}

%
Let us now analyze the interplay between the different contributions to $\ell_j\to \ell_i \ell_k \ell_k$ computed in Sec.~\ref{ssec:ndipole} 
and the correlations among LFV processes.
From the above discussion, it is clear that $\ell_j\to \ell_i \ell_k \ell_k$ is the dominant decay mode if the ALP can be produced on-shell, i.e.~for $2\,m_{\ell_i}< m_a < m_{\ell_j}-m_{\ell_i}$. On the other hand, this comparison is less evident below and above the resonant mass interval, 
as the interplay of the tree and loop-level contributions to $\ell_j\to 3\,\ell_i$ becomes non-trivial. 
To better illustrate this feature, in Fig.~\ref{fig:mue-illustration} we plot $\mathcal{B}(\ell_{j} \to 3\,\ell_i)/\mathcal{B}(\ell_j\to \ell_i\gamma)$ 
as a function of $m_a$ for $\mu\to e$ (left panel) and $\tau\to \mu$ (right panel) transitions. In this plot, the ALP couplings are set to $c_{\gamma\gamma}=1/(16\pi^2)$ and $a_{jj}^\ell=1$, with $a_{ii}^\ell/a_{jj}^\ell$ fixed to a few representative values. The LFV coupling $a_{ji}^\ell$ 
is not specified as it cancels in the ratios we are interested in, while the other couplings are neglected. 

\begin{figure}[t]
\centering
\includegraphics[width=0.5\textwidth]{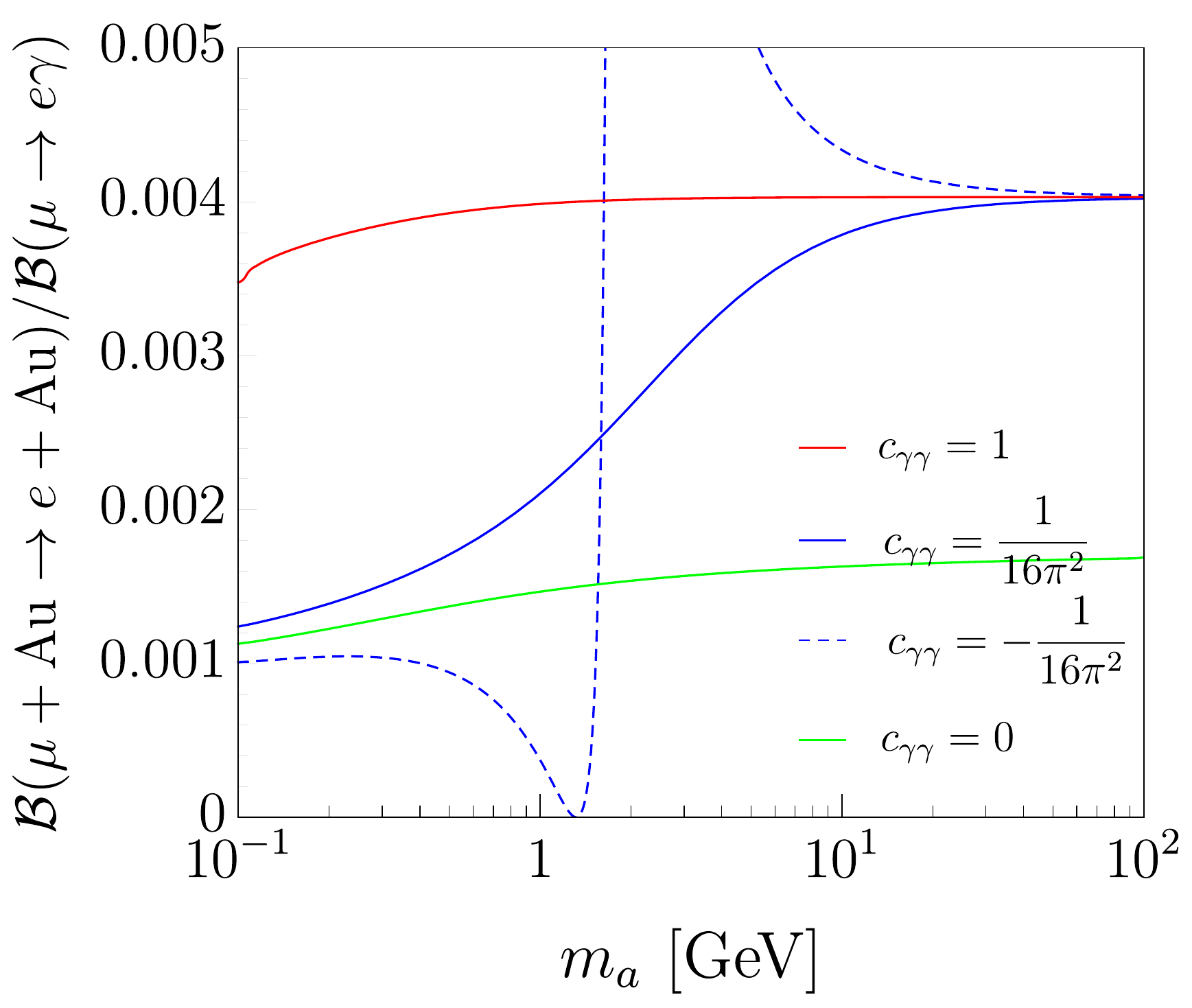}
\caption{\small\sl $\mathcal{B}(\mu +\! \mathrm{Au} \!\to\! e +\! \mathrm{Au})/\mathcal{B}(\mu \!\to\! e\gamma)$ 
is plotted as a function of $m_a$ for $a^\ell\equiv a_{ii}^\ell$ universal and various values of the coupling $c_{\gamma\gamma}$. We assume that $a_{\mu e}^\ell$ is the only non-vanishing LFV coupling.}
\label{fig:mueconv_mueg}
\end{figure}

From Fig.~\ref{fig:mue-illustration} we see that for $m_a >m_{\ell_j}$ the tree-level contribution to $\ell_j \to 3\,\ell_i$ decouples as $1/m^4_a$,
becoming subdominant compared to the loop-induced photon-penguin contributions, which have a milder logarithmic dependence on $m_a$. 
We find that tree-level contributions to $\mathcal{B}(\mu\to 3\,e)$ can be neglected for $m_a \gtrsim 1$~GeV, while masses as large as $m_a \approx 100$~GeV are needed  for $\tau\to 3\,\mu$ (see also Fig.~\ref{fig:mue-constraints}). In both cases, for sufficiently large $m_a$ values, $\mathcal{B}(\ell\to 3\,\ell_i)/\mathcal{B}(\ell_j\to \ell_i\gamma)$ approaches the prediction of dipole form-factor dominance, 
\begin{equation}
\label{eq:mu3e-mueg}
\dfrac{\mathcal{B}(\ell_j\to 3\ell_i)}{\mathcal{B}(\ell_j\to\ell_i \gamma)} \approx \dfrac{\alpha_\mathrm{em}}{3\pi}\left( \log \dfrac{m_{\ell_j}^2}{m_{\ell_i}^2}-\dfrac{11}{4}\right) \,,
\end{equation}
which yields $\approx 6\times 10^{-3}$, $\approx 10^{-2}$ and $\approx 2\times 10^{-3}$ for the $\mu\to e$, $\tau \to e$ and $\tau \to \mu$ transitions, respectively. In principle, this relation can be broken by contributions to $\ell_j\to 3\,\ell_i$ stemming from the anapole form-factors $\mathcal{F}_1(q^2)$ and $\mathcal{G}_1(q^2)$. Nonetheless, in most cases these contributions are sub-dominant with respect to the 
tree-level ones or the ones originating from dipoles. This can be understood from the ultraviolet logarithmic enhancement of the Barr-Zee dipole form-factor in~Eq.~\eqref{eq:F2} and \eqref{eq:G2}, and/or the kinematical logarithmic enhancement of Eq.~\eqref{eq:Gamma-loop}. Lastly, we also 
comment on the case where $m_a <2\, m_{\ell_i}$. For these values of $m_a$, the tree-level contribution dominates for $|a_{ii}^\ell|/|a_{ij}^\ell|\lesssim 1$, 
as depicted in Fig.~\ref{fig:mue-illustration}. Loop-induced contributions can become dominant only for large values of $a_{ii}^\ell$ compared to $a_{jj}^\ell$ 
and $c_{\gamma\gamma}$.

Another peculiar signature of ALP-induced LFV arises in the correlation between $\mu + N \to e + N$ and $\mu\to e \gamma$.
If the effects from dipole form-factors are dominant, as predicted in many NP scenarios, the following prediction holds
%
\begin{align}
\frac{\mathcal{B}(\mu^{-}+{\mathrm{Au}\rightarrow e^{-}{\mathrm{Au}}})}{\mathcal{B}(\mu\to e\gamma)} 
&\approx 4 \times 10^{-3} \,.
\label{eq:mueconv_mueg_num}
\end{align}
To investigate possible departures from this relation, we plot the same ratio in Fig.~\ref{fig:mueconv_mueg} for universal ALP couplings $a_{ii}^\ell \equiv a^\ell$ and various values of $c_{\gamma\gamma}$. For large values of $c_{\gamma\gamma}$ (red line), we find that the dipole effects driven by the Barr-Zee contributions dominate for any values of $m_a$, reproducing the results of Eq.~(\ref{eq:mueconv_mueg_num}). For $c_{\gamma\gamma}= \pm 1/16 \pi^2$, the above relation is broken by the anapole contributions for small values of $m_a$. Instead, for large $m_a$, the Barr-Zee effects dominate due to their logarithmic sensitivity to $m_a$, cf.~Eq.~\eqref{eq:F2}, and Eq.~\eqref{eq:mueconv_mueg_num} holds. Finally, for $c_{\gamma\gamma}=0$, anapole and dipole effects have a comparable size for any value of $m_a$ yielding a significant departure from Eq.~(\ref{eq:mueconv_mueg_num}).

Before moving into the discussion of processes involving hadrons, we turn our attention to the anomalous magnetic moment of leptons $(g-2)_\ell$ which are tightly related to LFV processes, as we are going to see in the following.

\subsection{On the $(g-2)_e$ and $(g-2)_\mu$ anomalies}

The anomalous magnetic moment of leptons, $a_{\ell} = (g-2)_\ell/2$, provides one of the most accurate tests of the SM validity. The longstanding discrepancy between the experimental value and the SM prediction, $\Delta a_\mu= a_\mu^\mathrm{exp}-a_\mu^{\mathrm{SM}}=(27.1\pm 7.3) \times 10^{-10}$ \cite{Bennett:2006fi,Keshavarzi:2018mgv,Davier:2017zfy}, at the level of $\approx 3.6\,\sigma$, received increased attention recently due to the anticipated new experimental results by the Muon $g-2$ Collaboration at Fermilab~\cite{Grange:2015fou}. Furthermore, the recent measurement of the fine-structure constant $\alpha_{\mathrm{em}}$ in atomic physics experiments~\cite{Parker:2018vye} allow us to concretely use for the first time the electron $g-2$ as a NP probe~\cite{Giudice:2012ms}. Surprisingly, the reevaluation of $\Delta a_e$ employing the latest value of $\alpha_{\mathrm{em}}$ shows a mild discrepancy $\Delta a_e=(-87\pm 36)\times 10^{-14}$~\cite{Hanneke:2008tm}, at the level of $2.4\sigma$. This value not only has a different sign than $\Delta a_\mu$, it also shows a departure from the expectations derived by considering the ``naive scaling", i.e.~the assumption that $\Delta a_\ell$ scales as $m_\ell^2$, which is valid for a large class of NP models~\cite{Giudice:2012ms}.

Light ALPs are promising candidates to accommodate both $g-2$ anomalies since they can contribute in different ways to $(g-2)_\ell$,  with $\ell=e,\mu,\tau$, breaking the ``naive scaling" expectations. The leading ALP contributions are given by the one-loop diagrams shown in Fig.~\ref{fig:diagram-mueg} (with $i=j$), as well as by the two-loop light-by-light and vacuum polarization diagrams, which are entirely induced by the $a\gamma\gamma$ coupling~\cite{Marciano:2016yhf}. The expression for $a_{\ell_i}=\mathcal{F}_{2}^{ii}(0)$ can then be written as
\begin{align}
\Delta a_{\ell_i} = \left( \Delta a_{\ell_i} \right)_{\rm LFC} + \left( \Delta a_{\ell_i} \right)_{\rm LFV}\,,
\label{eq:master_formula_gm2}
\end{align}
where the two contributions derive from lepton flavor-conserving (LFC) and LFV couplings, respectively.
Indeed, although $a_{\ell_i}$ are flavor-conserving observables, in presence of LFV couplings they can receive contributions from loops involving a lepton of different flavor, cf.~Fig.~\ref{fig:diagram-mueg} with $i=j \neq k$. The LFC expression from Yukawa interactions is universal and reads
\begin{align}
\left( \Delta a_{\ell_i} \right)_{\rm LFC} = -\frac{ m_{\ell_i}^{2}}{16 \pi^{2} \Lambda^{2}} &\Bigg{[} 
 64  \pi \,  \alpha_\mathrm{em} \, c_{\gamma \gamma}\, a_{ii}^\ell   \left(\log \frac{\Lambda ^2}{m_{\ell_i}^{2}} - h_{2}(x_i)\right)  +
 4 \, |a_{ii}^{\ell}|^2\, h_{1}(x_i) \Bigg{]}\,.
\label{eq:gm2_LFC}
\end{align}
Two-loop contributions to $\left( \Delta a_{\ell_i} \right)_{\rm LFC}$ induced by $a\gamma\gamma$ are almost negligible for the ALP couplings we consider~\cite{Marciano:2016yhf}.
The LFV contributions are different for each lepton flavor and they can be expressed as
\begin{align}
\left(\Delta a_{e}\right)_{\rm LFV} &=  \frac{ m_{e}^{2}}{16 \pi^{2} \Lambda^{2}} \Bigg{[} 
 \frac{m_{\mu} }{m_{e}} (|v_{e \mu}^{\ell}|^2-|a_{e \mu}^{\ell}|^2) \, g_{3}(x_\mu)
+  \frac{m_{\tau} }{m_{e}} (|v_{e \tau}^{\ell}|^2-|a_{e \tau}^{\ell}|^2) \, g_{3}(x_\tau)\Bigg{]} \,,
\nonumber\\[0.3em]
\left(\Delta a_{\mu}\right)_{\rm LFV}  &=  \frac{ m_{\mu}^{2}}{16 \pi^{2} \Lambda^{2}} \Bigg{[}  
 (|a_{e \mu}^{\ell}|^2 + |v_{e\mu}^{\ell}|^2) \,  h_{3}(x_\mu) + \frac{m_{\tau} }{m_{\mu}} (|v_{\mu \tau}^{\ell}|^2-|a_{\mu \tau}^{\ell}|^2) \, g_{3}(x_\tau)  \Bigg{]} \,,
\nonumber\\[0.3em]
\left(\Delta a_{\tau}\right)_{\rm LFV} &=  \frac{ m_{\tau}^{2}}{16 \pi^{2} \Lambda^{2}} \Bigg{[}   
(|a_{e \tau}^{\ell}|^2 + |v_{e\tau}^{\ell}|^2) \,  h_{3}(x_\tau) + (|a_{\mu \tau}^{\ell}|^2 + |v_{\mu \tau}^{\ell}|^2) \, h_{3}(x_\tau)\Bigg{]}\,,
\label{eq:gm2_LFV}
\end{align}
where $x_i=m_a^2/m_{\ell_i}^2$, as before, and the loop functions $h_i(x)$ and $g_i(x)$ are reported in Appendix A.  Note, in particular, that  $\left( \Delta a_{\ell_i} \right)_{\rm LFV}$ receives a chiral enhancement of order $m_{\tau}/m_{\ell_i}$ for $\ell_i=e,\mu$, hereby violating the ``naive scaling". We find that the leading UV-sensitive term in Eq.~\eqref{eq:gm2_LFC} agrees with~\cite{Marciano:2016yhf}, while the sub-leading finite terms proportional to $h_{1,2}(x_i)$ agree with Refs.~\cite{Leveille:1977rc} and \cite{Bauer:2019gfk}, respectively. Moreover, the LFV contributions depicted in Eq.~\eqref{eq:gm2_LFV} agree with the recent results from Ref.~\cite{Bauer:2019gfk}. 

The inspection of Eqs.~(\ref{eq:master_formula_gm2}) and (\ref{eq:gm2_LFV}) leads to the following remarks:

\begin{itemize}
\item The functions $h_1(x)$, $h_2(x)$ and $g_3(x)$ are identically positive, while the sign of $h_3(x)$ depends on the value of the 
ALP mass~\cite{Bauer:2019gfk}. Therefore, LFC contributions from Yukawa interactions (cf.~right panel of Fig.~\ref{fig:diagram-mueg})
cannot account for the $(g-2)_\mu$ anomaly due to the wrong sign, as noted before in Ref.~\cite{Leveille:1977rc,Giudice:2012ms,Marciano:2016yhf,Bauer:2019gfk}.
\item Flavor conserving contributions of Barr-Zee type (cf.~left panel of Fig.~\ref{fig:diagram-mueg}) can accommodate both $(g-2)_e$ and $(g-2)_\mu$ anomalies provided $|c_{\gamma\gamma}|\gtrsim 10\times|a_{\mu\mu}^\ell|$ and $|a_{ee}^\ell| \gtrsim 10 \times |a_{\mu\mu}^\ell|$, with $c_{\gamma\gamma}\, a_{\mu\mu}^\ell < 0$ and $a_{ee}^\ell\, a_{\mu\mu}^\ell < 0$. This is illustrated in Fig.~\ref{fig:gm2_FC} for two benchmark values for $m_a$, namely $m_a=1$~GeV and $10$~GeV, and for fixed values of $a_{ee}^\ell/\Lambda = 10 \,\mathrm{TeV}^{-1}$ (left panel) and $50 \,\mathrm{TeV}^{-1}$ (right panel). Although phenomenologically viable, this scenario requires large couplings to electrons and photons, which might be challenging to obtain in a ultraviolet complete scenario. In this plot, the different shape of the $(g-2)_{e,\mu}$ constraints can be traced back to the interplay of Barr-Zee and pure Yukawa effects, cf.~Eq.~(\ref{eq:master_formula_gm2}). For a sufficiently large (small) $a_{\mu\mu}^\ell$ ($c_{\gamma\gamma}$) coupling, the Yukawa contribution -- which has the wrong sign to explain the $(g-2)_{\mu}$ anomaly -- tends to dominate over the Barr-Zee one, setting a lower bound on $c_{\gamma\gamma}$. This lower bound is relaxed for increasing values of $m_a$ as the Yukawa effects decouple faster than the Barr-Zee ones.

\begin{figure}[t]
\centering
\includegraphics[width=0.45\textwidth]{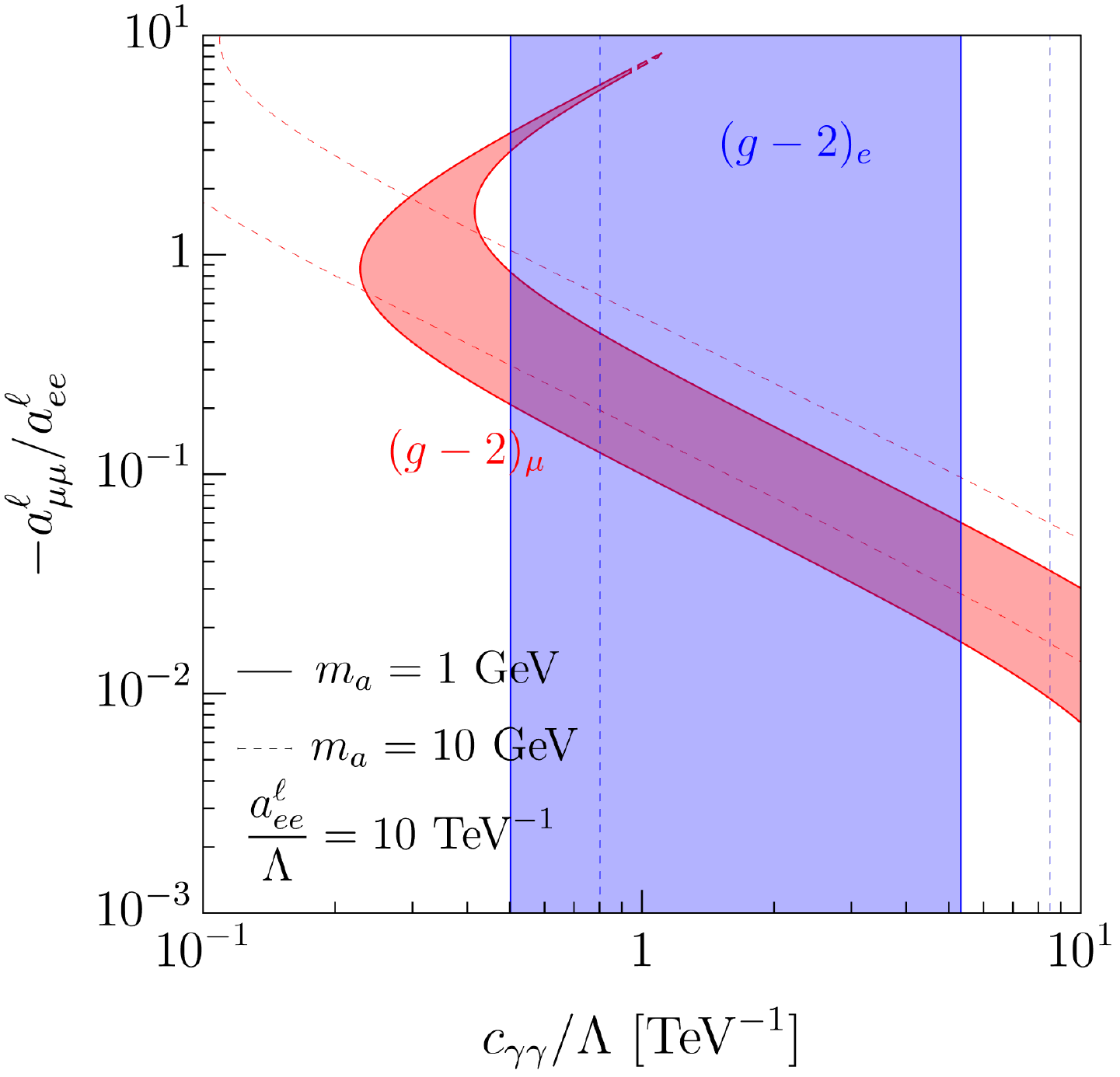}~~~~~~~~\includegraphics[width=0.45\textwidth]{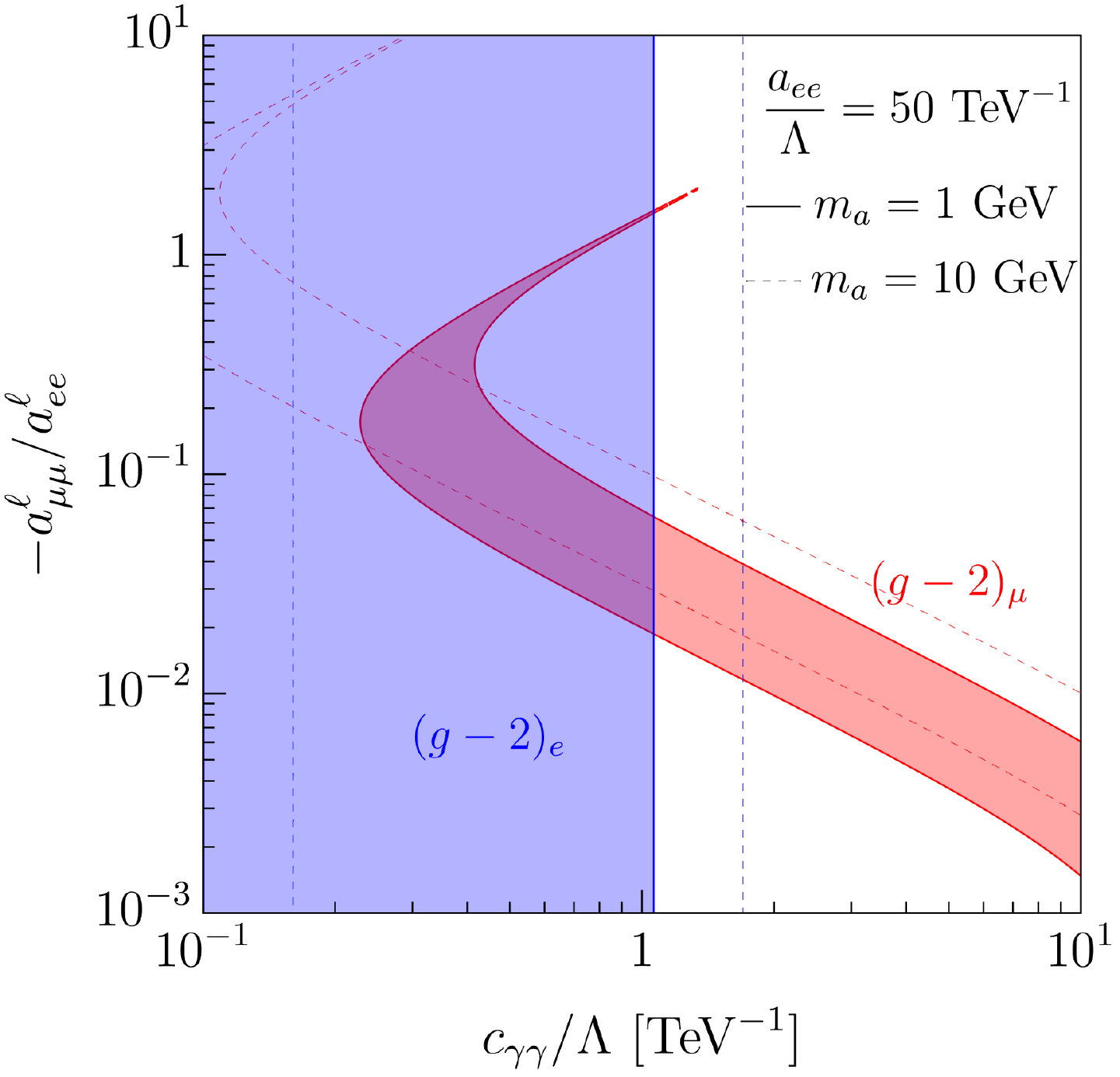}
\caption{\small\sl 
Allowed regions in the $-a_{\mu \mu}^\ell/a_{ee}^\ell$ vs. $c_{\gamma \gamma}/\Lambda$ plane where the $(g-2)_{\mu}$ and $(g-2)_{e}$ anomalies
are accounted for at the $95\%$ C.L.. The plots have been obtained setting $\Lambda = 1$ TeV, 
$a_{ee}^\ell/\Lambda = 10 \,\mathrm{TeV}^{-1}$ (left) and $a_{ee}^\ell/\Lambda = 50 \,\mathrm{TeV}^{-1}$ (right). In each plot we consider two benchmarks for the ALP mass, namely $m_{a} = 1 \, \, \mathrm{GeV}$ (solid lines) and $m_{a}=10 \,\, \mathrm{GeV}$ (dashed lines). 
}
\label{fig:gm2_FC}
\end{figure}

%
\item LFV contributions can be especially relevant for $(g-2)_e$ given the large $m_\tau/m_e$ chiral enhancement in Eq.~\eqref{eq:gm2_LFV}. A viable explanation of this discrepancy can be easily obtained via LFV contributions induced by the coupling $a_{e\tau}^\ell$. Such scenario would remain viable as long as $|a^\ell_{\mu\tau}/ a^\ell_{e\tau}|\lesssim 10^{-4}$ and $|a^\ell_{\tau\tau}/ a^\ell_{e\tau}|\lesssim 10^{2}$,
in order to avoid the experimental bounds from $\mu\to e\gamma$ and $\tau\to e\gamma$, respectively.~\footnote{
In principle, LFV contributions can accommodate both $(g-2)_e$ and $(g-2)_\mu$ anomalies through a unique $a_{e\mu}^\ell$ coupling~\cite{Bauer:2019gfk}. However, we stress that such a solution would require a huge hierarchy 
$| a^\ell_{\mu\mu}/a^\ell_{e\mu}|\lesssim 10^{-7}$ in order to avoid the experimental bound on $\mu\to e\gamma$, 
making this possibility less compelling.} It is natural to wonder if this mechanism could be combined with the Barr-Zee solution to accomodate the $(g-2)_\mu$ discrepancy as well. For this combined explanation to work, the ALP mass must satisfy $m_a > m_\tau$, in order to evade stringent constraints from $\tau \to e\mu\mu$, which would be otherwise resonantly enhanced by tree-level ALP contributions. This scenario is illustrated in Fig.~\ref{fig:gm2_LFV}, by fixing $m_a=5$~GeV and $c_{\gamma\gamma}/\Lambda = 0.1~\mathrm{TeV}^{-1}$. From this plot, we see that the main constraint to a simultaneous explanation of $(g-2)_e$ and $(g-2)_\mu$ comes from $\tau\to e \gamma$, which strongly depends on the product of couplings $a_{\tau e}^\ell \, c_{\gamma\gamma}$, where $a_{\tau e}^\ell$ and $c_{\gamma\gamma}\, a_{\mu\mu}^\ell$ are fixed by $(g-2)_e$ and $(g-2)_\mu$, respectively. There is a small window in which both discrepancies can be explained, but this solution requires considerably large values of $a_{\mu\mu}^\ell$. This corner of parameter space is expected to be fully probed by future $\tau\to e\gamma$ searches at Belle-II, as shown in Fig.~\ref{fig:gm2_LFV}.

\begin{figure}[t]
\centering
\includegraphics[width=0.45\textwidth]{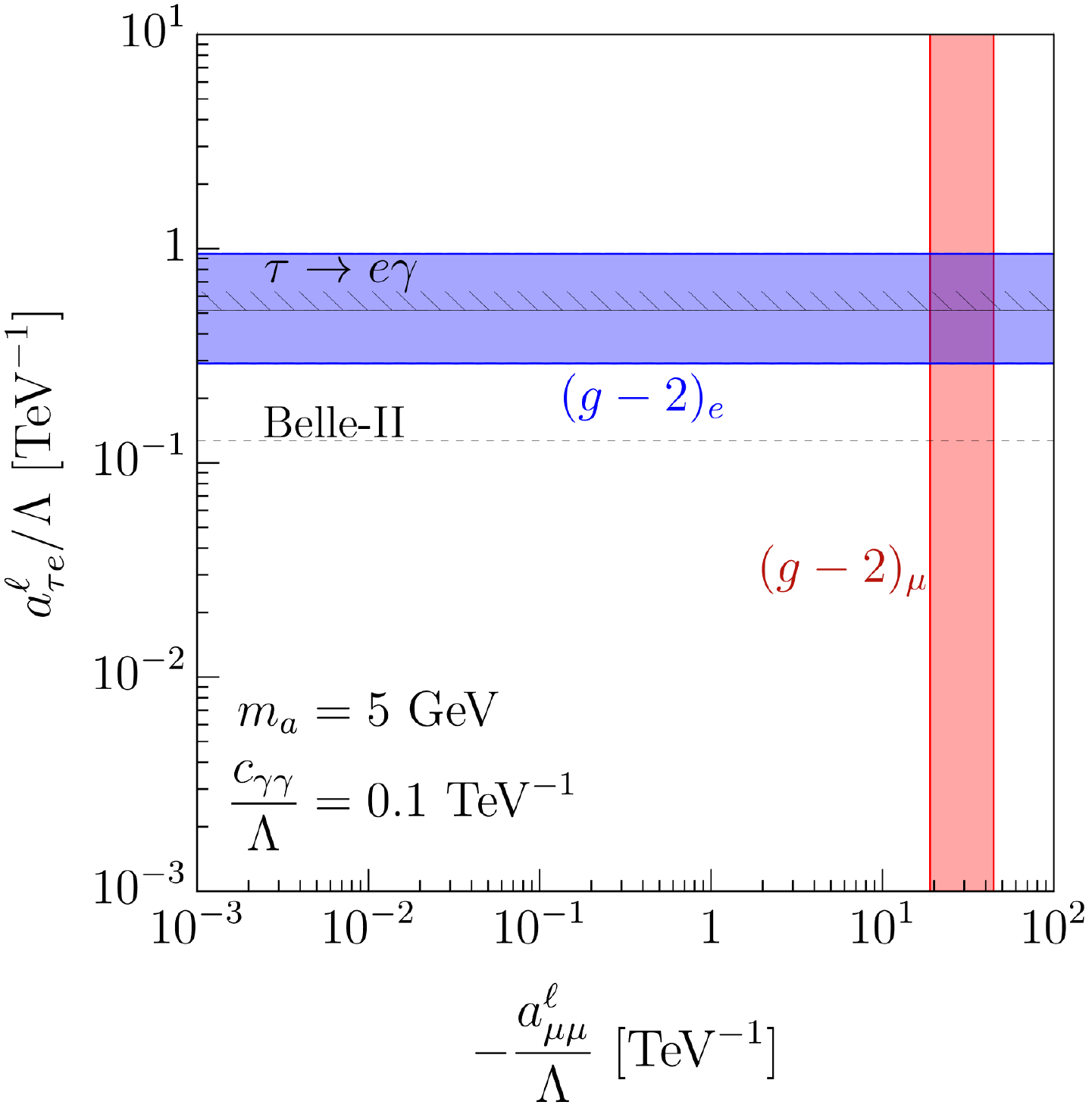}
\caption{\small\sl 
Allowed regions in the $a_{\tau e}^\ell/\Lambda$ vs.~$-a_{\mu\mu}^\ell/\Lambda$ plane where the $(g-2)_{\mu}$ and $(g-2)_{e}$ anomalies
are accounted for at the $95\%$ C.L. The plot has been obtained setting $\Lambda = 1$ TeV, $c_{\gamma\gamma}/\Lambda = 0.1 \,\mathrm{TeV}^{-1}$ and $m_{a} = 5 \, \,\mathrm{GeV}$. 
For simplicity, the other ALP couplings have been set to zero. Regions excluded by present (future) limits on $\mathcal{B}(\tau \to e \gamma)$ are shown by the solid (dashed) black lines, cf.~Table~\ref{tab:exp-lep}.}
\label{fig:gm2_LFV}
\end{figure}

%
\item For completeness, we also comment on $(g-2)_\tau$. In this case, there is no parametric enhancement for LFV effects, since $m_\tau$ 
is the largest fermionic mass, therefore the dominant effects stem from LFC couplings. For $a_{\tau\tau}, v_{\tau\tau} \sim \mathcal{O}(1)$, 
we expect that $|\Delta a_\tau| \lesssim 10^{-5}$, still far from the poor sensitivity of current experiments~\cite{Eidelman:2007sb}.
\end{itemize}


\section{Hadronic processes}
\label{sec:had}

LFV decays of mesons also provide a powerful probe of ALP interactions. On the one hand, they are highly complementary to the purely leptonic processes discussed in Sec.~\ref{sec:lep} as they are sensitive to different combinations of ALP couplings. On the other hand, we expect significant experimental improvements in the coming years thanks to the effort of the NA62~\cite{Petrov:2017wza}, LHCb~\cite{Bediaga:2018lhg,Borsato:2018tcz} 
and Belle-II~\cite{Kou:2018nap,Cerri:2018ypt} collaborations.

There are three types of hadronic processes that can be studied experimentally: (i) two body-decays $P\to \ell_i \ell_j$, (ii) semileptonic decays $P\to P^{\prime} \ell_i\ell_j$ and (iii) $P\to V \ell_i\ell_j$, where $P^{(\prime)}$ and $V$ denote generic pseudoscalar and vector mesons, respectively. 
Concerning the two body-decays, we focus on pseudoscalar mesons instead of vector ones, since parity conservation implies that $\langle 0 | G_{\mu\nu} \widetilde{G}^{\mu\nu}| V \rangle =0$, while the relation $\langle 0 | \bar{q}^{(\prime)} \gamma_5 q | V\rangle=0$ can be derived from the Ward identities.
We will derive now the most general expressions for the branching ratios of these processes and discuss their potential to probe ALP couplings. Our notation is such that leptonic charged conjugated final states are added $\ell_i \ell_j \equiv \ell_i^\pm \ell_j^\mp= \ell_i^- \ell_j^+ + \ell_i^+ \ell_j^-$ and therefore
$\mathcal{B}(P \to (M) \ell_i \ell_j) = 2\,\mathcal{B}(P \to (M)\ell_i^- \ell_j^+)$ where $M=P^{\prime}, V$.

\subsection{$P \to \ell_i \ell_j$ and $\ell_j \to \ell_i P$}

\subsubsection{General expressions}

\begin{figure}[h!]
\centering
\includegraphics[width=0.65\textwidth]{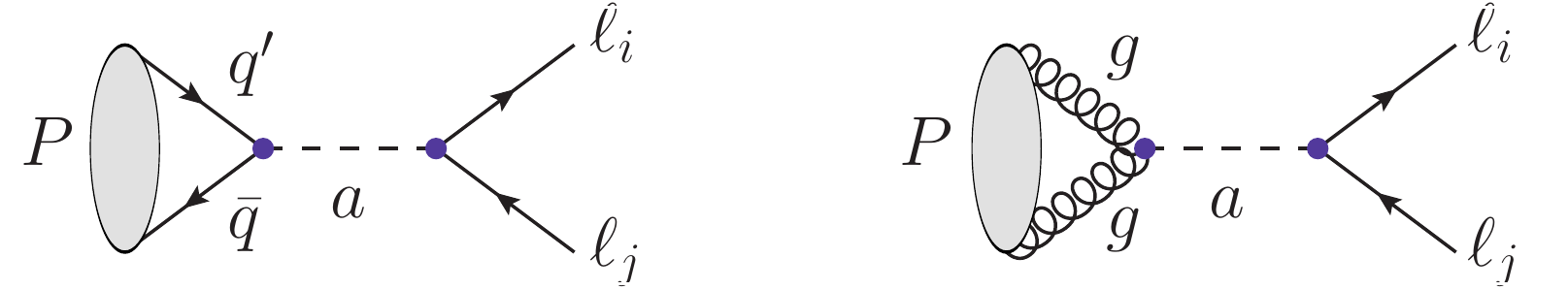}
\caption{\sl \small Diagrams contributing to the processes $P\to \ell_i \ell_j$ and $\ell_{j} \to \ell_{i} P$ via the ALP couplings to quarks (left panel) and gluons 
(right panel), where $P$ denotes a generic pseudoscalar meson. The latter diagram only contributes to processes involving light unflavored mesons.}
\label{fig:diagram-hadronic}
\end{figure}

We start by considering the simplest LFV hadronic processes, namely the decays $P \to \ell_i \ell_j$. These processes can be induced via ALP 
couplings to quarks and/or gluons, as illustrated in Fig.~\ref{fig:diagram-hadronic}. While the quark contribution is always present, the gluonic one can only contribute to decays of unflavored mesons, such as $P= \pi, \eta, \eta^\prime$.

Assuming that $m_{\ell_j}\gg m_{\ell_i}$, the branching fraction for $P\to \ell_i \ell_j$ is given by
\begin{align}
\label{eq:Plilj}
\begin{split}
\mathcal{B}(P \to \ell_i \ell_j) = ~
&\tau_P\,\dfrac{m_P\,m_{\ell_j}^2}{4\pi\Lambda^4} \dfrac{|\mathcal{N}_P|^2}{(m_P^2-m_a^2)^2+m_a^2 \Gamma_a^2}\,\left(1 - \dfrac{m_{\ell_j}^2}{m_P^2} \right)^{\! 2} |s_{ij}^\ell|^2\,,
\end{split}
\end{align}

\noindent where $\mathcal{N}_P \equiv\Lambda \, \langle 0 | \mathcal{L}_{\mathrm{eff}}^{D \leq 5}| P \rangle$ is a function of the ALP 
couplings to quarks and gluons, and of the relevant hadronic parameters, which can be expressed as follows
\begin{align}
\label{eq:NP}
\begin{split}
\mathcal{N_P}= & ~16\pi^2 \,c_{gg} \,a_P -i a_{ij}^q(m_{q_i}+m_{q_j}) \langle 0| \bar{q}_i\gamma_5 q_j |P \rangle \,,
\end{split}
\end{align}
\noindent where the summation over quark-flavor indices is implicit and the hadronic constant $a_P$ is defined as 
\begin{align}
\label{eq:def-ap}
a_{P} \equiv \langle 0 | \frac{\alpha_s}{4\pi} G_{\mu\nu} \widetilde{G}^{\mu\nu}| P \rangle  \,.
\end{align}
\noindent In Eq.~\eqref{eq:NP}, the matrix element $\langle 0| \bar{q_i}\gamma_5 q_j |P \rangle$ depends on the relevant meson decay 
constants which should be determined together with $a_P$ by non-perturbative means, as will be discussed below.

If kinematically allowed, the same couplings generating $P \to \ell_i \ell_j$ also generate the inverse process $\ell_j \to P\ell_i$. The corresponding branching fraction is
\begin{equation}
\begin{split}
\mathcal{B}(\ell_j \to P \ell_i) = ~&\tau_{\ell_i}\dfrac{m_{\ell_j}^3}{16\pi\Lambda^4} \dfrac{|\mathcal{N}_P|^2}{(m_P^2-m_a^2)^2+m_a^2 \,\Gamma_a^2}\,\left( 1 - \dfrac{m_P^2}{m_{\ell_j}^2} \right)^2 |s_{ij}^\ell|^2\,
\end{split}\,,
\end{equation}

\noindent which depends on the parameter $\mathcal{N}_P$ defined above. 

\subsubsection{Hadronic inputs: $\mathcal{N}_P$}

The next step is to derive the expression for $\mathcal{N}_P$, defined in Eq.~\eqref{eq:NP}, for each pseudoscalar meson. This quantity depends on the pseudoscalar density, $\langle 0 |q_i \gamma_5 q_j | P \rangle$, as well as on the anomaly matrix element, parameterized by $a_P$ in Eq.~\eqref{eq:def-ap}. 

\paragraph{Flavored mesons} We start by considering heavy flavored mesons such a $D^0$ and $B_{(s)}$. In this case, the only non-vanishing contribution comes from the second term in Eq.~\eqref{eq:NP}. By expressing the heavy-light meson as $P=\overline{Q}q$, the axial matrix element reads
\begin{equation}
\langle 0 |\bar{Q} \gamma^\mu \gamma_5 q | P(p) \rangle = i f_P p^\mu\,,
\end{equation}
from which one can show that 
\begin{equation}
\langle 0 |\bar{Q} \gamma_5 q | P \rangle = -i \dfrac{m_P^2 \,f_P}{m_q+m_Q}\,,
\end{equation}
where $f_P$ is the $P$-meson decay constant. By replacing this expression in Eq.~\eqref{eq:NP}, we obtain
\begin{align}
\mathcal{N}_{D}&= - a^u_{12}\, f_{D}\, m_{D}^2\,,\\
\mathcal{N}_{B_d}&=- a^d_{13}\, f_{B}\, m_{B_d}^2\,,\\
\mathcal{N}_{B_s}&=- a^d_{23}\, f_{B_s}\, m_{B_s}^2\,.
\end{align}
In these expressions, the only needed inputs are $f_P$, which have been determined in all cases by means of numerical simulations of QCD on the lattice, cf.~Table~\ref{tab:hadronic}. Similarly, in the kaon system, we define $\vert K_{L(S)} \rangle = (\vert K^0 \rangle \pm \vert \overline{K^0} \rangle )/\sqrt{2}$ and write
\begin{align}
\mathcal{N}_{K_L} &= -\sqrt{2} \,\mathrm{Re} \left[ a_{12}^d \right] m_{K^0}^2 f_{K}\,, \\
\mathcal{N}_{K_S} &= -i \sqrt{2} \,\mathrm{Im} \left[ a_{12}^d \right] m_{K^0}^2 f_{K}\,,
\end{align}
\noindent showing that $K_{L(S)}$ leptonic decays can probe either the real or imaginary part of $a_{12}^d$.

\paragraph{Pseudoscalar quarkonia states} For the heavy quarkonia states $\eta_c$ and $\eta_b$, we find in a similar way that
\begin{align}
\mathcal{N}_{\eta_c} &= 16 \pi^2 c_{gg}\, a_{\eta_c}- a^u_{22}\, f_{\eta_c} m_{\eta_c}^2\,,\\
\mathcal{N}_{\eta_b} &= 16 \pi^2 c_{gg}\, a_{\eta_b}- a^d_{33}\, f_{\eta_b} m_{\eta_b}^2\,,
\end{align}

\noindent where $f_{\eta_c}$ and $f_{\eta_b}$ are also listed in Table~\ref{tab:hadronic}. The anomaly contribution is 
also present in this case, but it is sub-dominant since these particles are much heavier than $\Lambda_{\mathrm{QCD}}$.

\begin{table}[!t]
\centering
  \renewcommand{\arraystretch}{1.5} 
\begin{tabular}{|c|cc|}
\hline
Quantity & Value [MeV] & Ref. \\
\hline\hline
$f_{\pi}$ & $130.2(0.8)$ & \cite{Aoki:2016frl} \\
$f_{K}$ & $155.7(0.3)$ & \cite{Aoki:2016frl} \\
$f_{D}$ & $212.0(0.7)$ & \cite{Aoki:2016frl} \\
$f_{B_d}$ & $190.0(1.3)$ & \cite{Aoki:2016frl} \\
$f_{B_s}$ & $230.3(1.3)$ & \cite{Aoki:2016frl} \\
\hline
$f_{\eta_c}$ & $391(4)$ & \cite{Becirevic:2013bsa}  \\
$f_{\eta_b}$ & $667(6)$  & \cite{McNeile:2012qf} \\ \hline
\end{tabular}
\caption{\em \small Decay constants computed by means of numerical simulations of QCD on the lattice.}  
\label{tab:hadronic}
\end{table}

\paragraph{Light unflavored mesons}

Finally, we discuss the more subtle case of $\pi^0$ and $\eta^{(\prime)}$ mesons. For pions, one can use the exact isospin limit to derive the pseudoscalar density, while the anomaly contribution can be obtained by taking the divergence of the axial current~\cite{Gross:1979ur,Brignole:2004ah}. We find that

\begin{equation}
\label{eq:Npi}
\mathcal{N}_\pi = - \dfrac{f_\pi\,m_\pi^2}{\sqrt{2}} \left(a^u_{11}-a^d_{11}\right)-{16  \pi^2\, \dfrac{1-z}{1+z}\, \dfrac{f_\pi\, m_\pi^2}{\sqrt{2}}\, c_{gg}}\,,
\end{equation}

\noindent where $z=m_u/m_d$, in such a way that the anomaly contribution vanishes in the isospin conserving limit, $m_u=m_d$. For $\eta^{(\prime)}$, the anomaly contribution plays an even more important role. By denoting $q=u,d$ and taking $m_q = (m_u+m_d)/2$, the pseudoscalar densities can be parameterized as 
\begin{align}
2 m_q \, \langle 0 \vert \bar{q} \gamma_5 q \vert \eta^{(\prime)} \rangle &= -\dfrac{i}{\sqrt{2}}\,h_{\eta^{(\prime)}}^q\,,\\
2 m_s \, \langle 0 \vert \bar{s} \gamma_5 s \vert \eta^{(\prime)} \rangle &= -i\,h_{\eta^{(\prime)}}^s\,,
\end{align}

\noindent where $h_{\eta^{(\prime)}}^q$ and $h_{\eta^{(\prime)}}^s$ are decay constants.~\footnote{Note that the axial and pseudoscalar densities are not directly related to the same decay constant for $\eta$ and $\eta^\prime$ since the anomalous contribution is relevant in this case.}  These definitions allow us to write
\begin{align}
\label{eq:Neta}
\begin{split}
\mathcal{N}_{\eta^{(\prime)}}=16\pi^2 c_{gg} \,a_{\eta^{(\prime)}} &- (a^u_{11}+a^d_{11})\,\dfrac{h_{\eta^{(\prime)}}^q}{\sqrt{2}}-a^d_{22}\,h_{\eta^{(\prime)}}^s\,.
\end{split}
\end{align}

\noindent The best available computation of $a_{\eta^{(\prime)}}$, $h_{\eta^{(\prime)}}^{q}$ and $h_{\eta^{(\prime)}}^{s}$ relies on the so-called Feldmann-Kroll-Stech (FKS) mixing scheme~\cite{Feldmann:1998vh}. This phenomenological approach is based on the assumption that the states $\vert\eta_q\rangle = (\vert u\bar{u}+\vert d\bar{d}\rangle)/\sqrt{2}$ and $\vert\eta_s \rangle = \vert s\bar{s}\rangle$ only mix through the anomaly. In this case, by using inputs such as the mixing angle between $\eta$ and $\eta^\prime$, the authors of Ref.~\cite{Feldmann:1998vh,Beneke:2002jn} obtained the inputs collected in Table~\ref{tab:eta-etaprime}.

\begin{table}[!h]
\centering
  \renewcommand{\arraystretch}{1.7} 
\begin{tabular}{|c|ccc|}
\hline
$P$ & $h_P^q$ $[\mathrm{GeV}^3]$ & $h_P^s$ $[\mathrm{GeV}^3]$ & $a_P$ $[\mathrm{GeV}^3]$  \\
\hline\hline
$\eta$ & $0.001(3)$ & $-0.055(3)$ & $-0.022(2)$\\ 
$\eta^\prime$ & $0.001(2)$ & $0.068(5)$ & $-0.057(2)$\\ \hline
\end{tabular}
\caption{\em \small Hadronic inputs for $\eta$ and $\eta^\prime$ obtained in Ref.~\cite{Feldmann:1998vh,Beneke:2002jn} by using the FKS mixing scheme.}  
\label{tab:eta-etaprime}
\end{table}

\subsection{$P\to P^\prime \ell_i\ell_j$}

\subsubsection{General expressions}

The next processes we consider are semileptonic decays of the type $P\to P^\prime \ell_i\ell_j$. These processes can be induced via the ALP couplings 
to quarks, cf.~Fig.~\ref{fig:diagram-hadronic} with $q\neq q^\prime$, and are complementary to the processes $P\to \ell_i\ell_j$ described above due to the parity symmetry. More specifically, since these decays arise from the underlying transition $q_k \to q_l \ell_i^- \ell_j^+$ (with $i\neq j$ and $k\neq l$), 
the relevant hadronic matrix elements read

\begin{equation}
\langle P^\prime(k) \vert \bar{q_l} \gamma_5 q_k \vert P(p)\rangle =0\,,
\qquad\qquad
\langle P^\prime(k) \vert \bar{q_l} q_k \vert P(p) \rangle = f_0 (q^2) ~ \dfrac{m_P^2-m_{P^\prime}^2}{m_{q_k}-m_{q_l}}\,,
\end{equation}

\noindent where $q^2=(p-k)^2$ and $f_0(q^2)\equiv f_0^{P\to P^\prime}(q^2)$ denotes the $P\to P^\prime$ scalar form factor. Since the pseudoscalar matrix element vanishes, these decays can only constrain the vector ALP couplings $v_{kl}^q$. The general branching fraction is then given by~\cite{Becirevic:2016zri}

\begin{equation}
\dfrac{\mathrm{d} \mathcal{B}}{\mathrm{d}q^2}(P \!\to\! P^\prime \ell_j \ell_i) = \dfrac{\tau_P m_P  m_{\ell_j}^2}{64 \pi^3}
\dfrac{ q^2 \lambda_{P^\prime}^{1/2} f_0(q^2)^2}{(q^2 \!- m_a^2)^2+ m_a^2\, \Gamma_a^2}\left(1-\dfrac{m_{\ell_j}^2}{q^2}\right)^{\!\!2}
\!\!
\left(1-\dfrac{m_{P^\prime}^2}{m_P^2}\right)^{\!\!2}  \dfrac{|v_{kl}^q|^2 |s_{ij}^\ell|^2}{\Lambda^4}\,,
\end{equation}

\noindent where we have assumed once again $m_{\ell_j}\gg m_{\ell_i}$, and the phase-space function is given by $\lambda_{P^\prime} \equiv \lambda(m_P,m_{P^\prime},\sqrt{q^2})$. This expression can be directly applied to the decays $D\to \pi$, $D_s \to K$ and $B\to K$, 
among others. The only subtle case regards kaon decays, for which the above expression should be amended by replacing $|v_{kl}^q|^2$ by
\begin{alignat}{2}
K^+ &\to \pi^+: \qquad && |v_{21}^d|^2\,,\\
K_L &\to \pi^0: && \mathrm{Im}[v_{21}^d]^2\,,\\
K_S &\to \pi^0: && \mathrm{Re}[v_{21}^d]^2\,.
\end{alignat}

\noindent In other words, the different neutral kaon decays can probe either the real or imaginary parts of the Wilson coefficients, depending on their $CP$ properties. 

The largest contribution to $\mathcal{B}(P\to P^\prime \ell_i \ell_j)$ arises when the ALP can be produced on-shell, i.e.~for $m_a\in (m_{\ell_i}+m_{\ell_j},m_P-m_{P^\prime})$. In this case, the above expression can be simplified by means of the narrow-width approximation

\begin{equation}
\mathcal{B}(P\to P^\prime a \to P^\prime \ell_i \ell_j) = \mathcal{B}(P\to P^\prime a)\,\mathcal{B}(a\to \ell_i \ell_j)\,,
\end{equation}
\noindent where

\begin{equation}
\mathcal{B}(P\to P^\prime a)= \dfrac{\tau_P\, m_P\,f_0(m_a^2)^2}{16\pi} \left(1-\dfrac{m_{P^\prime}^2}{m_P^2}\right)^2 \dfrac{|v_{kl}^q|^2}{\Lambda^2}\lambda^{1/2}(m_P,m_P^\prime,m_a)\,,
\end{equation}
and the leptonic branching fraction is given in Eq.~\eqref{eq:alilj}.~\footnote{Note that our notation is such that $\mathcal{B}(a\to \ell_i \ell_j) \equiv \mathcal{B}(a\to \ell_i^\pm \ell_j^\mp) = 2\,\mathcal{B}(a\to \ell_i^- \ell_j^+)$.}

\subsubsection{Hadronic inputs}

The last theoretical input needed in the above expression is the form factor $f_0$. The latest LQCD results are summarized in Ref.~\cite{Aoki:2019cca}. For the $B\to \pi$ and $B\to K$ transitions, for which LQCD form factors are not available in the full $q^2$-range, we use results from light-cone sum rules~\cite{Ball:2004ye}. 

\subsection{$P\to V \ell_i\ell_j$}

\subsubsection{General expressions}

We now turn to $P\to V$ semileptonic decays, with $V$ being a generic vector meson, which turn out to be complementary to the observables described above.  In this case, the relevant matrix elements are
\begin{align}
\langle V(k) \vert \bar{q_l} \gamma_5 q_k \vert P(p)\rangle &=-i \left(\varepsilon\cdot q\right)\dfrac{2\, m_V}{m_b+m_s}A_0(q^2)\,, & \langle V(k) \vert \bar{q_l} q_k \vert P(p) \rangle &= 0\,,
\end{align}

\noindent where $q^2=(p-k)^2$, $\varepsilon^\mu$ denotes the $V$-meson polarization, and $A_0\equiv A_0^{P\to V}$ stands for the pseudoscalar 
$P\to V$ form factor. Since the scalar matrix element vanishes in this case, these decays can only probe the axial coupling $a_{kl}^q$, differently from 
the $P\to P^\prime$ processes described above, that are only sensitive to $v_{kl}^q$. The general expression for their branching fraction can be recast 
from Ref.~\cite{Becirevic:2016zri}, giving

\begin{align}
\begin{split}
\dfrac{\mathrm{d} \mathcal{B}}{\mathrm{d}q^2}(P &\to V \ell_i \ell_j) = \dfrac{\tau_P\, m_{\ell_j}^2}{64 \pi^3\, m_P^3}\,\dfrac{|a_{kl}^q|^2 |s_{ij}^\ell|^2}{\Lambda^4}\left(1-\dfrac{m_{\ell_j}^2}{q^2}\right)^2\dfrac{ q^2\, \lambda_{V}^{3/2}(q^2) A_0(q^2)^2}{(q^2-m_a^2)^2+ m_a^2 \, \Gamma_a^2}\,,
\end{split}
\end{align}
\noindent where $\lambda_V(q^2)= \lambda(m_P,m_V,\sqrt{q^2})$. 

For on-shell ALP production, i.e.~$m_a\in (m_{\ell_i}+m_{\ell_j},m_P-m_V)$, the above formula can be simplified as

\begin{equation}
\mathcal{B}(P\to V a \to V \ell_i \ell_j) = \mathcal{B}(P\to V a) \, \mathcal{B}(a\to \ell_i \ell_j)\,,
\end{equation}

\noindent where the ALP branching fraction is given in Eq.~\eqref{eq:alilj} and
\begin{equation}
\mathcal{B}(P\to V a)= \dfrac{\tau_P\,A_0(m_a^2)^2}{16\pi\,m_P^3} \dfrac{|a_{kl}^q|^2}{\Lambda^2}\lambda_V^{3/2}(m_a^2)\,.
\end{equation}

\subsubsection{Hadronic inputs}

The most relevant $P\to V$ transition for our study is $B\to K^\ast$. Since the nedeed form factor is not yet available from LQCD simulations in the full $q^2$ range, we consider the combination of LQCD and light-cone sum rules results from Ref.~\cite{Straub:2015ica}.

\subsection{Numerical results and discussion}

In this section we derive the limits from existing data and compare the sensitivity of the different decay channels listed above. The present experimental constraints are collected in Table~\ref{tab:exp} along with the future prospects when available. 

Let us first consider the ALP couplings to gluons, $c_{agg}$, which can trigger LFV processes involving light unflavored mesons, namely $\pi^0$, $\eta$ and $\eta^\prime$. Starting with the $\mu \to e$ transition, the only kinematically-allowed processes are $\pi^0 \to \mu e$ and $\eta^{(\prime)} \to \mu e$. By combining Eq.~\eqref{eq:Plilj} with the hadronic inputs in Eq.~\eqref{eq:Npi} and \eqref{eq:Neta}, and neglecting the ALP couplings to quarks, we find that
\begin{align}
\label{eq:Pmue-1}
\mathcal{B}(P \to \mu e) =c_P^{\mu e}\, |s_{12}^\ell|^2\, \left(\dfrac{c_{gg}}{1/16\pi^2}\right)^2\, \left( \dfrac{\Lambda}{1\,\mathrm{TeV}}\right)^{-4}\, \left(\dfrac{m_{P}^2}{m_{P}^2-m_a^2}\right)^2\,, 
\end{align}
with 
\begin{align}
\label{eq:Pmue-2}
\lbrace c_{\pi^0}^{\mu e},\, c_\eta^{\mu e},\, c_{\eta^\prime}^{\mu e}\, \rbrace \simeq \lbrace 2.6,\, 0.02,\, 0.2 \rbrace \times 10^{-13}\,.
\end{align}
\noindent From these expressions, given the current experimental sensitivity depicted in Table~\ref{tab:exp}, as well as the existing limits on $s_{12}^\ell$ derived from leptonic observables in Sec.~\ref{sec:lep}, we conclude that such processes are not promising probes of ALPs. The only exception are the very narrow regions around $m_a \approx m_P$, where a resonant contribution is produced. Similar conclusions can be obtained by considering, instead of $c_{gg}$, the axial ALP couplings to quarks with appropriate flavor indices. 

\begin{figure}[t!]
\centering
\includegraphics[width=0.5\textwidth]{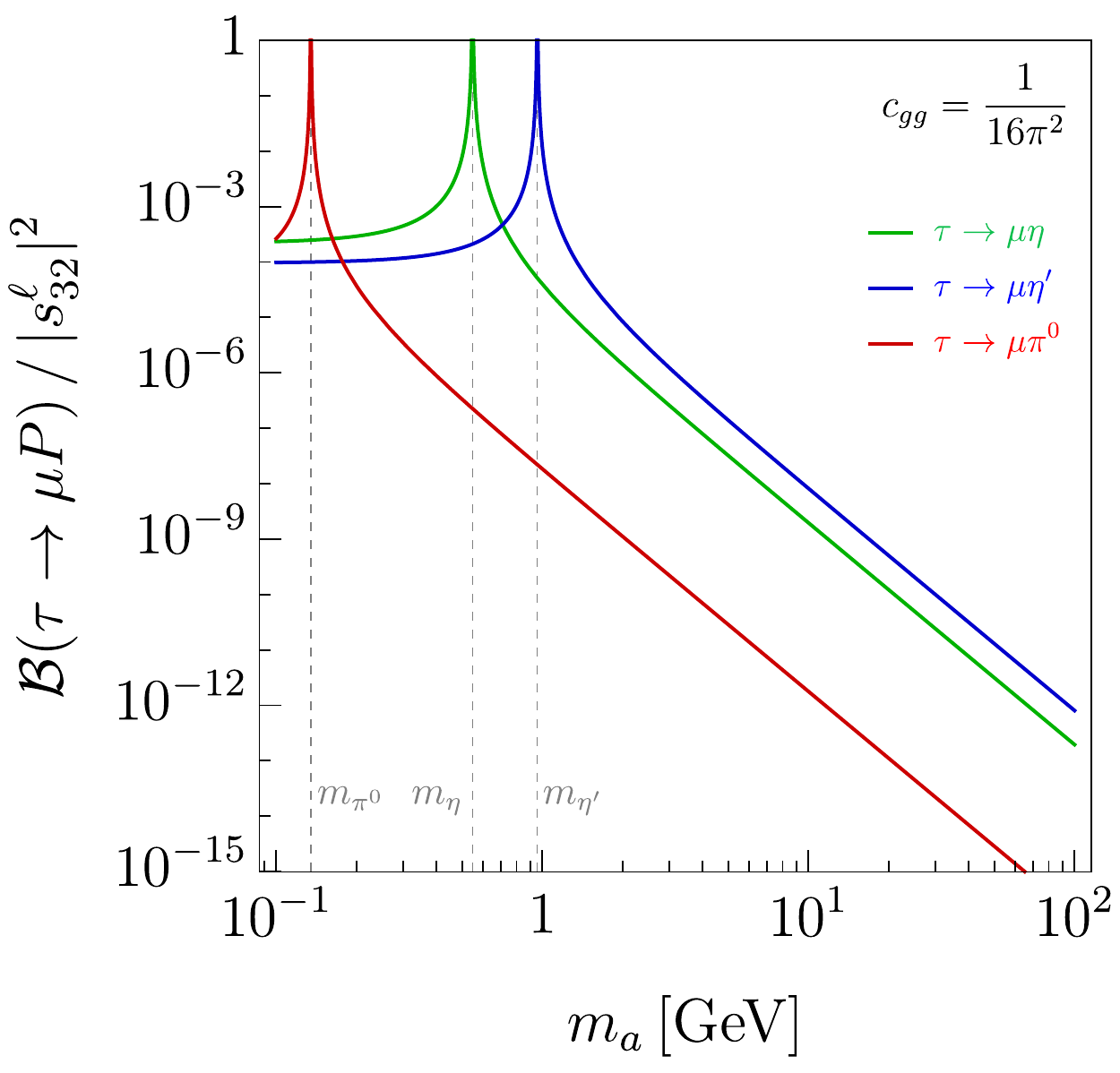}~\includegraphics[width=0.5\textwidth]{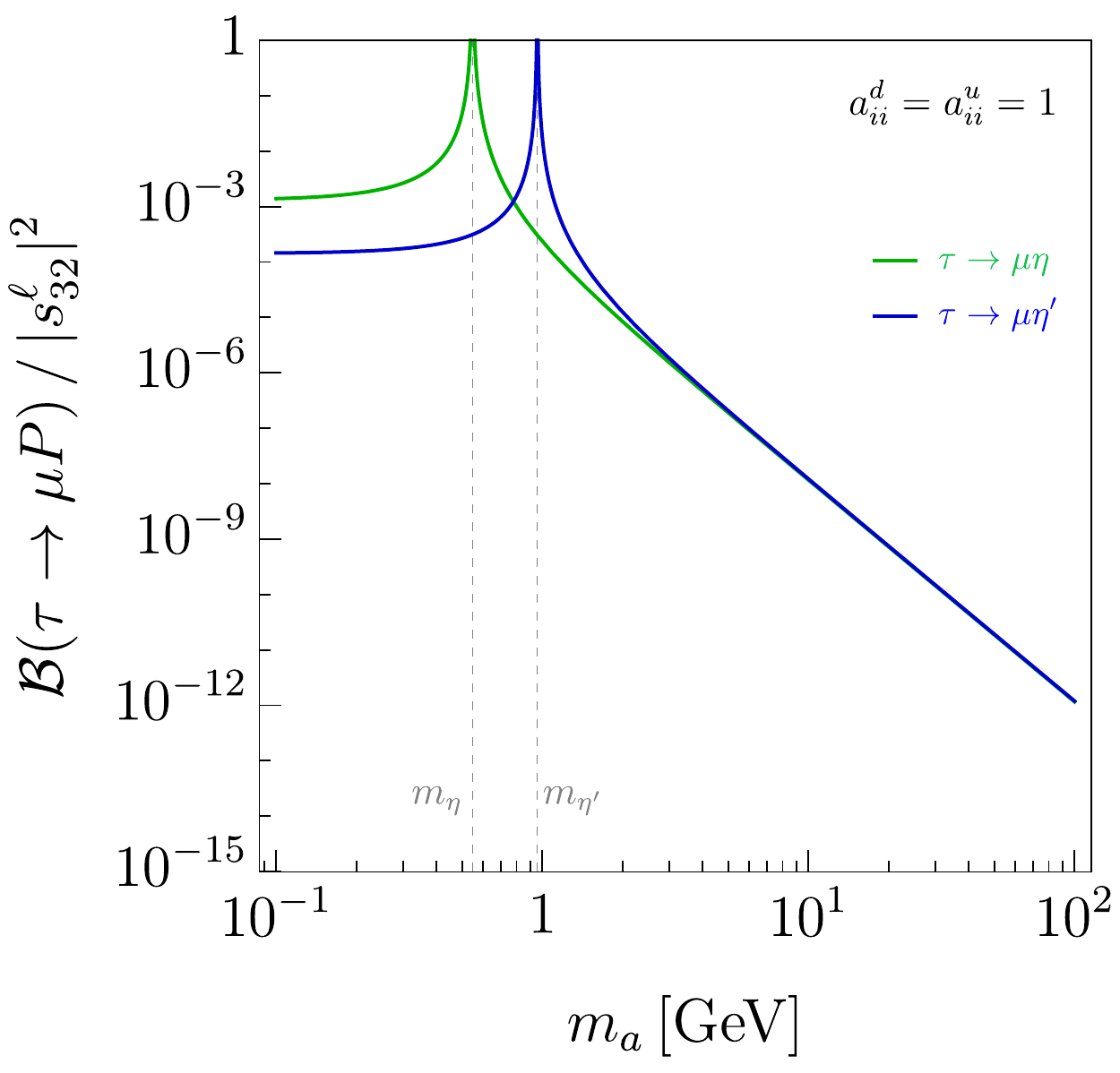}
\caption{\small\sl Normalized branching fraction $\mathcal{B}(\tau\to\mu P) / |s_{23}^\ell |^{2}$ plotted against $m_a$ for $P=\pi^0$ 
and $\eta^{(\prime)}$ in two benchmark scenarios, namely $c_{gg}=1/(16\pi^2)$ (left panel) and $a_{ii}^d=a_{ii}^u=1$ (right panel).}
\label{fig:tauPmu-pred}
\end{figure}

The situation is much more promising for the $\tau \to e$ and $\tau \to \mu$ transitions. In this case, the available processes are $\tau \to l \pi^0$ and $\tau \to l \eta^{(\prime)}$, with $l=e,\mu$. By only keeping the contributions driven by $c_{agg}$ and by focusing on the decays into muons, we find that
\begin{align}
\mathcal{B}(\tau \to \mu P) &\approx c_P^{\mu\tau}\, |s_{23}^\ell|^2\, \left(\dfrac{c_{gg}}{1/16\pi^2}\right)^2\, \left( \dfrac{\Lambda}{1\,\mathrm{TeV}}\right)^{-4}\,\left(\dfrac{m_{P}^2}{m_{P}^2-m_a^2}\right)^2\, \,,\\
\end{align}
with
\begin{align}
\lbrace c_{\pi^0}^{\mu \tau},\, c_\eta^{\mu \tau},\, c_{\eta^\prime}^{\mu \tau}\, \rbrace \simeq \lbrace 0.05,\, 2.0, 1.0 \rbrace \times 10^{-4}\,,
\end{align}
\noindent which are considerably larger than the values found in Eq.~\eqref{eq:Pmue-2}. These decay modes are particularly interesting 
given the expected experimental resolutions at Belle-II, which are going to improve the present limits by at least one order 
of magnitude \cite{Kou:2018nap}, cf.~Table \ref{tab:exp}. 
To understand why the processes $\tau \to \ell P$ are more sensitive to new physics than $P \to \mu e$, one should compare the total lifetime 
of the decaying particles. More precisely, one finds that 
\begin{equation}
\dfrac{\tau_\tau}{\tau_\pi} \approx 3.4\times 10^3\,,\qquad\dfrac{\tau_\tau}{\tau_\eta} \approx 5.8\times 10^5\,,\qquad\text{and}\qquad \dfrac{\tau_\tau}{\tau_{\eta^\prime}} \approx 8.6\times 10^7\,
\end{equation}
which can be understood from the fact that $\tau$'s can only decay through weak interactions, differently than $\pi^0$ and $\eta^{(\prime)}$. 
For that reason, $\tau$ decays are much better probes of new physics than light-meson decays. 

To further explore the potential of $\tau$ decays to constrain ALPs, in Fig.~\ref{fig:tauPmu-pred} we plot $\mathcal{B}(\tau \to \mu P)$, normalized to $|s_{23}^\ell|^2$, as a function of $m_a$ in two scenarios: (i) $c_{agg}=1/(16\pi^2)$ and (ii) $a_{ii}^d = a_{ii}^u=1$, with the other couplings taken to be zero. From these plots, we see that the largest branching fractions are obtained at the resonances, i.e.~$m_a \approx m_P$. However, large branching fractions can also be attained for other masses. Notice, also, that these decays modes have a complementary sensitivity to $c_{agg}$, $a_{ii}^d$ and $a_{ii}^u$, as it can be seen by comparing the left and right panels of Fig.~\ref{fig:tauPmu-pred}.

\begin{figure}[p!]
\centering
\includegraphics[width=0.5\textwidth]{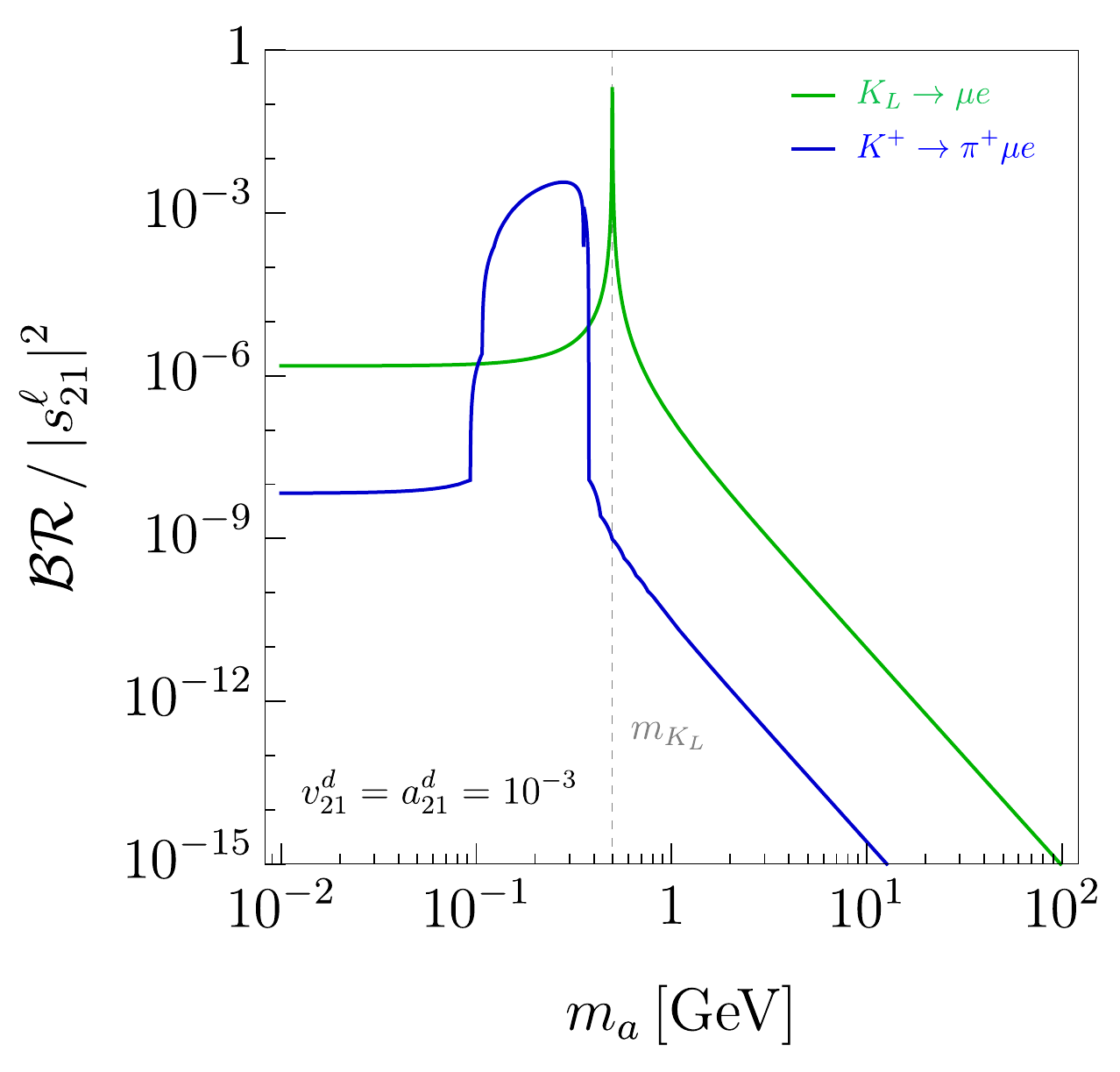}
\caption{\small\sl $\mathcal{B}(K^+\to \pi^+ \mu e)/|s^\ell_{21}|^2$ (blue line) and $\mathcal{B}(K_L\to \mu e)/|s^\ell_{12}|^2$ (green line) are plotted against $m_a$. The ALP couplings to quarks are taken to be $v_{21}^d=a_{21}^d=10^{-3}$, while the ALP width is fixed to $\Gamma_a=10^{-6}~\mathrm{GeV}$ for illustration. Different values of $\Gamma_a$ would imply a shift of the semileptonic rate in the resonant region.}
\label{fig:kaon-pred}
\end{figure}

\begin{figure}[p!]
\centering
\includegraphics[width=0.5\textwidth]{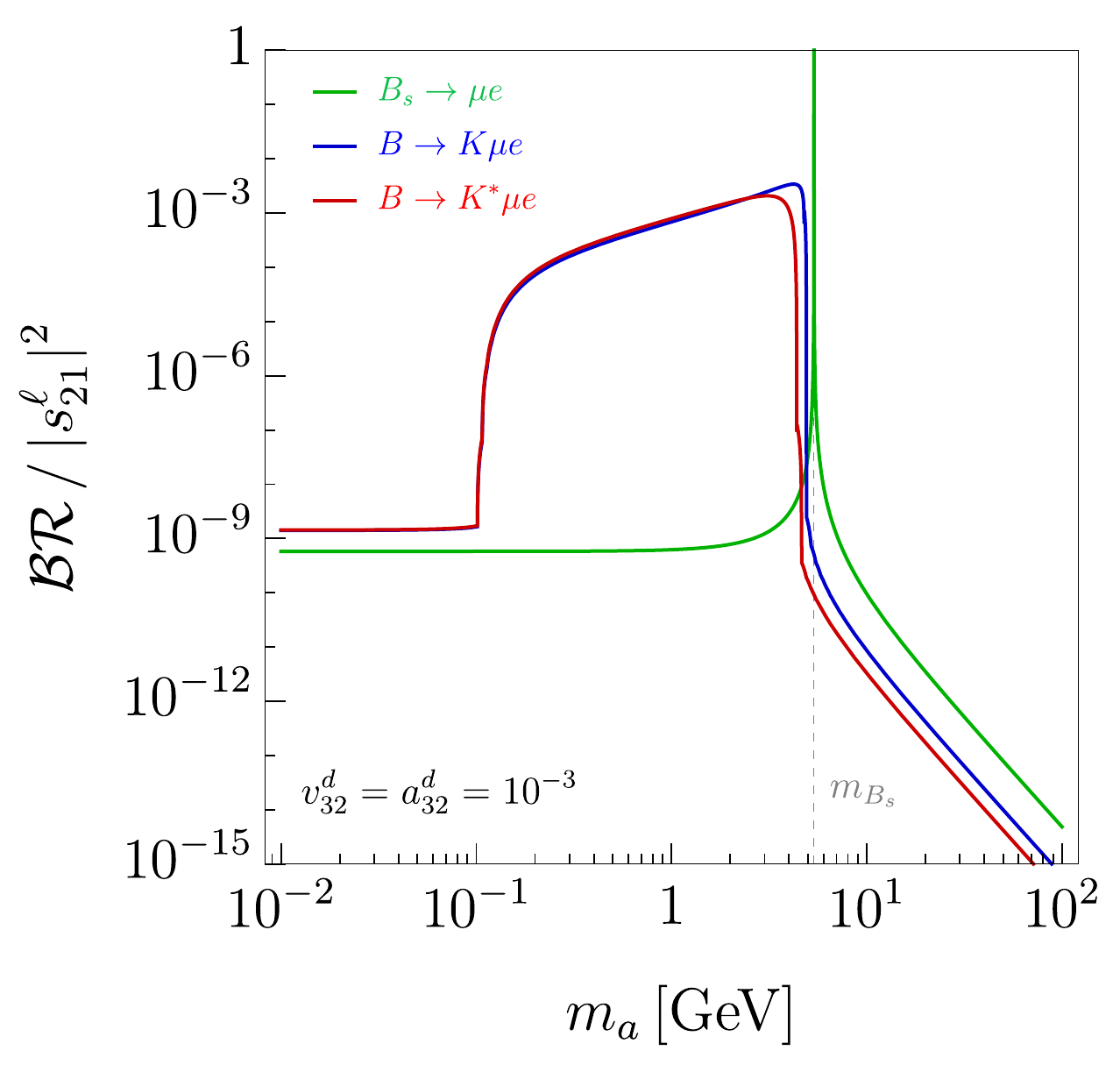}~\includegraphics[width=0.5\textwidth]{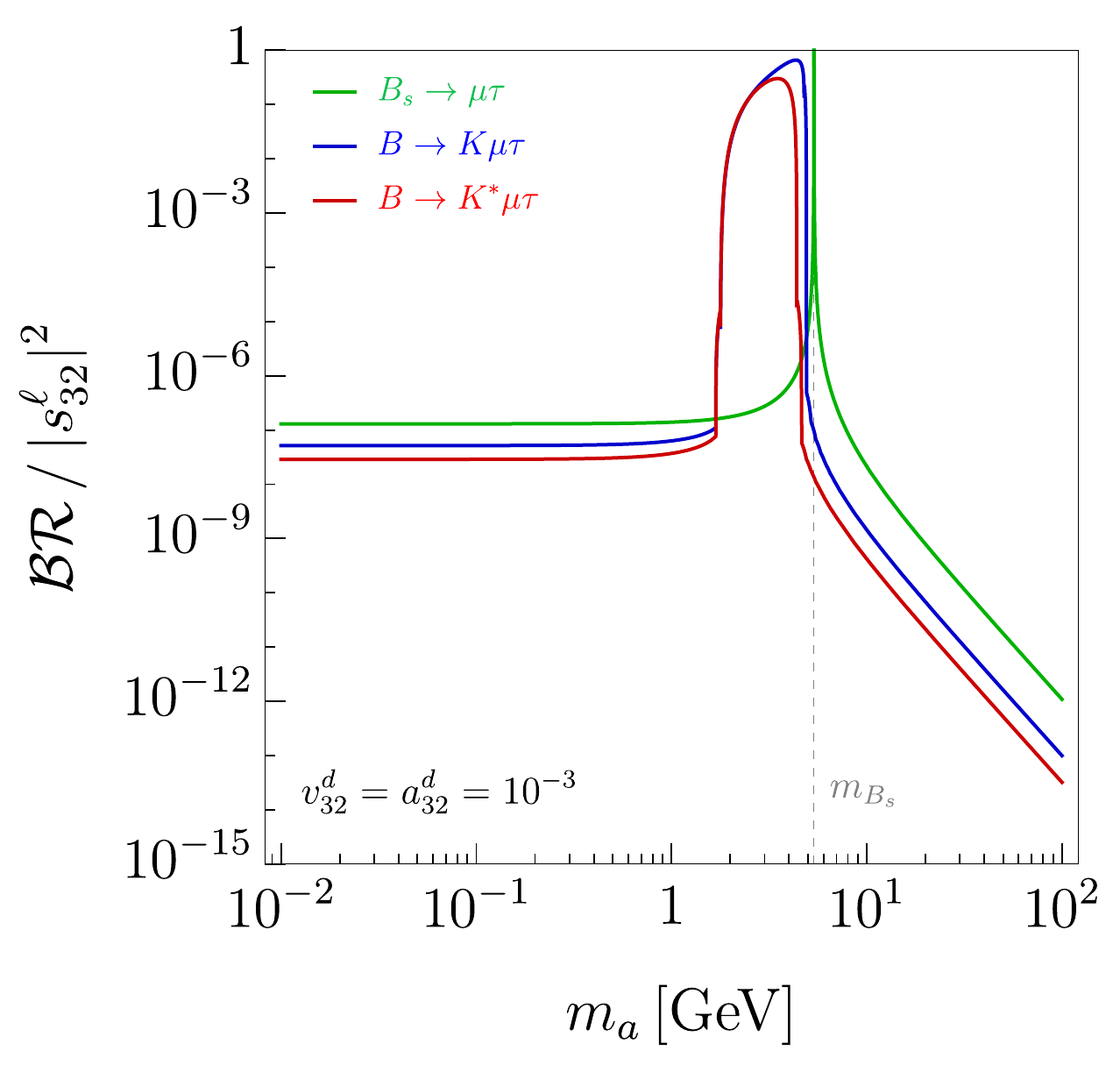}
\caption{\small\sl $\mathcal{B}(B\to K \ell_i \ell_j)/|s^\ell_{ij}|^2$ (green line), $\mathcal{B}(B\to K^\ast \ell_i \ell_j)/|s^\ell_{ij}|^2$ (blue line) and $\mathcal{B}(B_s\to \ell_i\ell_j)/|s^\ell_{ij}|^2$ (red line) are plotted against $m_a$ for $(i,j)=(1,2)$ (left panel) and $(2,3)$ (right panel). The ALP couplings to quarks are taken to be $v_{32}^d=a_{32}^d=10^{-3}$, while the ALP width is fixed to $\Gamma_a=10^{-6}~\mathrm{GeV}$ for illustration. }
\label{fig:BKll-pred}
\end{figure}

Next, we discuss the potential of (semi)leptonic kaon and $B$-meson decays to probe ALP couplings. These decay modes are very sensitive to NP contributions due to their quark-flavor changing nature. Furthermore, there is a rich experimental program at NA62, LHCb and Belle-II experiments that will improve the experimental sensitivity on many observables. Unlike the processes discussed above, (semi)leptonic decays can only probe the flavor violating ALP couplings to quarks, $a_{ij}^q$ and $v_{ij}^q$, with $i\neq j$. Leptonic decays are sensitive to axial couplings, while semileptonic decays can probe either the axial or vector ones, depending on the spin of the meson in the final state. For this reason, these processes provide complementary information on the NP couplings. 
To quantitatively compare the different decay modes, we assume that $a_{ij}^d=v_{ij}^d=10^{-3}$ and plot the branching fractions normalized by the LFV coupling as a function of $m_a$. This is shown in Fig.~\ref{fig:kaon-pred} and \ref{fig:BKll-pred} for kaon and $B_{(s)}$-meson observables, respectively. We find that the semileptonic rates are always the largest ones in the resonant regions, while the leptonic ones are typically dominant for large and/or small values of $m_a$, outside the resonant region.

\begin{figure}[t!]
\centering
\includegraphics[width=0.5\textwidth]{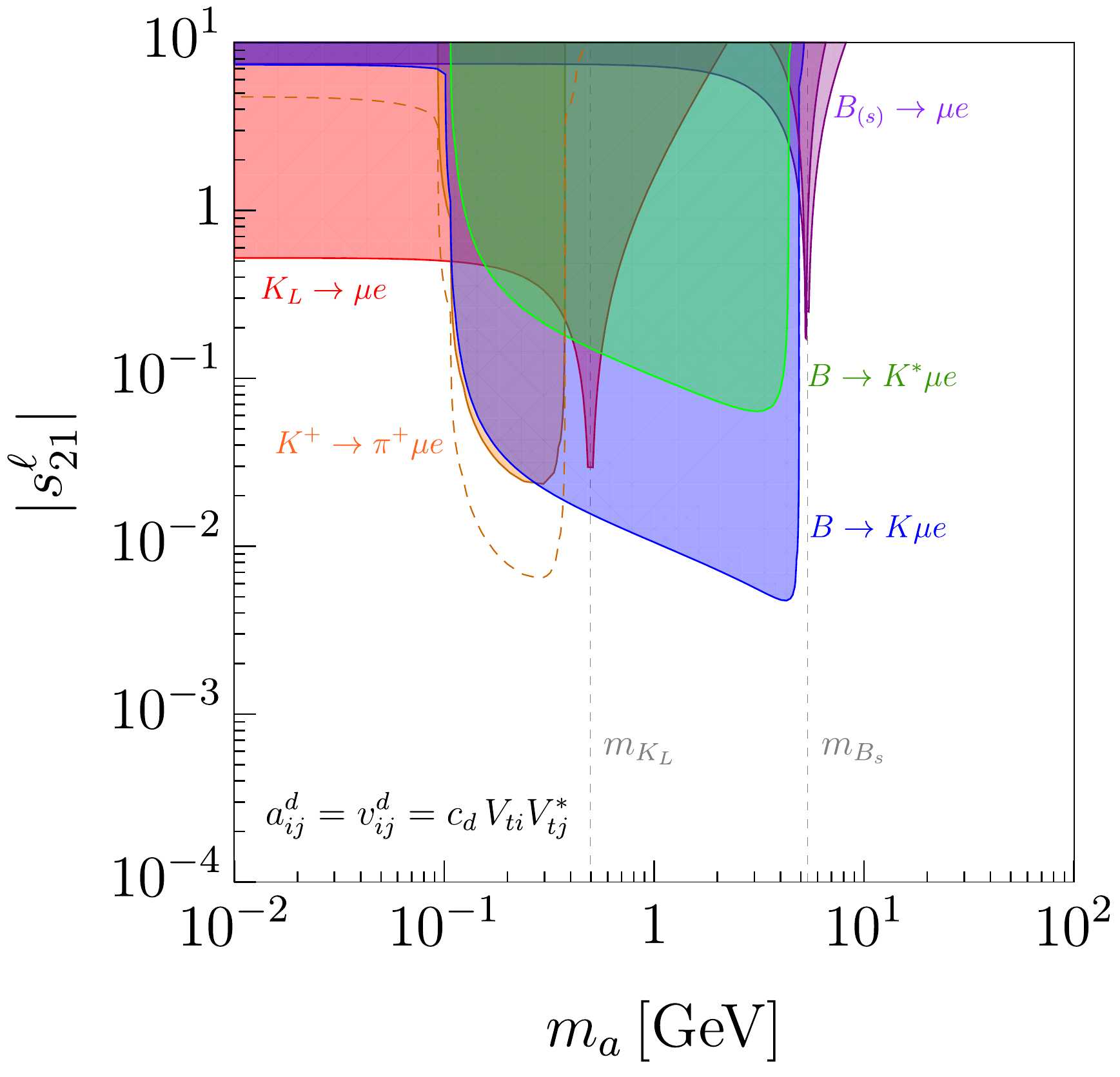}~\includegraphics[width=0.5\textwidth]{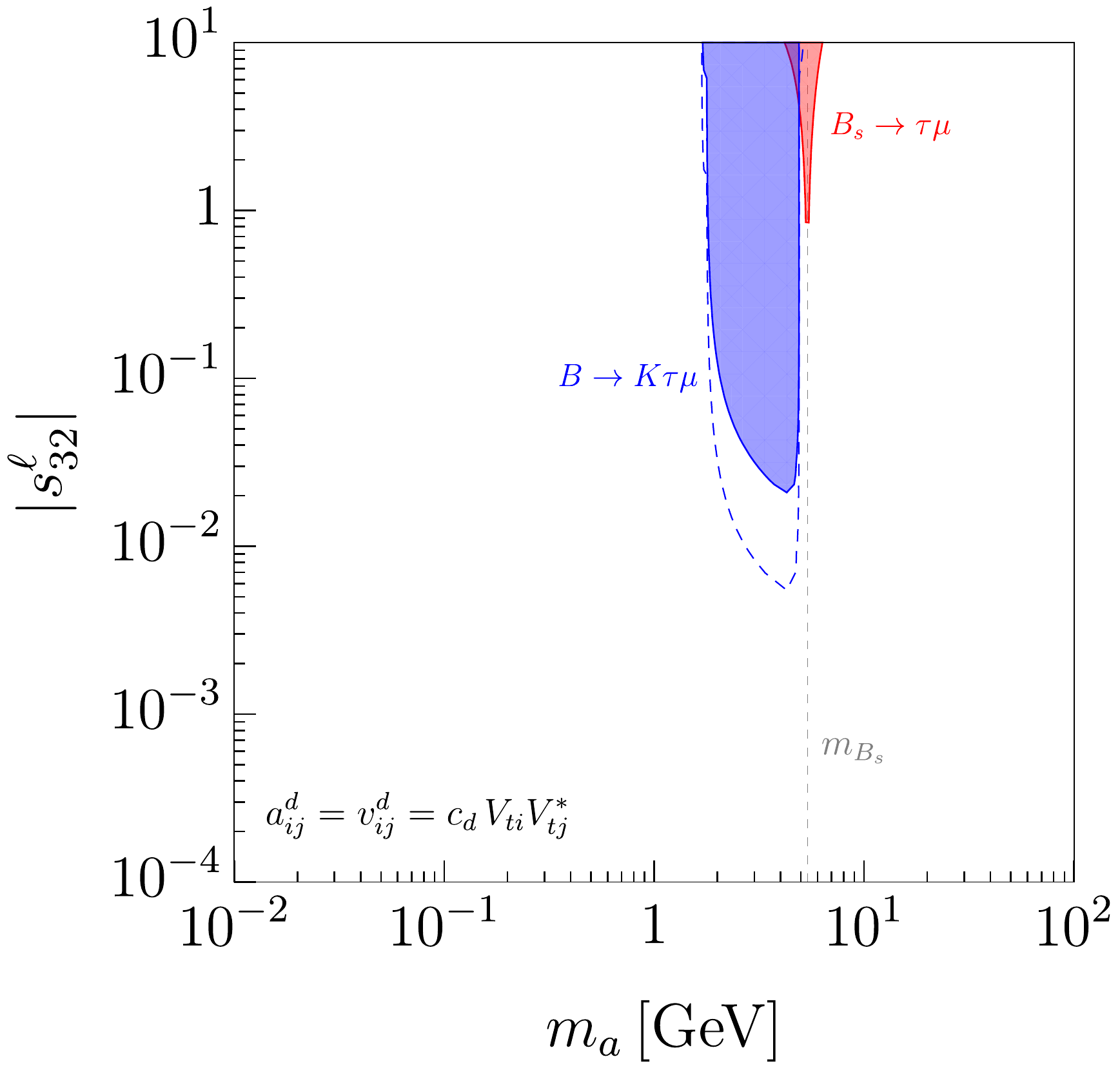}
\caption{\small\sl Constraints on $|s_{2 1}^\ell|$ (left panel) and $|s_{3 2}^\ell|$ (right panel) as a function of the ALP mass obtained from the hadronic probes listed in Table~\ref{tab:exp}. Dashed lines correspond to future experimental prospects. A benchmark scenario where $a_{ij}^d=v_{ij}^d \simeq c_d \,V_{ti} V_{tj}^\ast$, with $c_d \approx 10^{-2}$, has been considered, while the ALP width is fixed to an illustrative value $\Gamma_a = 10^{-6}$~GeV. Larger (smaller) values of $\Gamma_a$ would decrease (increase) the excluded regions in the on-shell regions for semileptonic decays. Constraints on $|s^\ell_{31}|$ are not depicted, since they turn out to be very similar to $|s^\ell_{32}|$.}
\label{fig:hadronic-constraints}
\end{figure}

To compare the sensitivity of kaon and $B$-meson decays channels, we consider a benchmark model that allows us to connect different quark transitions. We assume that the matrices $a^d$ and $v^d$ satisfy
\begin{equation}
a^d_{ij}=v^d_{ij} \simeq c_d\, V_{ti} V_{tj}^\ast\,,
\end{equation}
where $V$ denotes the CKM matrix and $c_d$ is a constant. Such relation is predicted, for instance, from electroweak loops in models with predominant ALP couplings to the $W$-boson and/or the top-quark~\cite{Gavela:2019wzg}. The constraints on $s^\ell_{ij}$ derived from kaon and $B$-meson decays are shown in Fig.~\ref{fig:hadronic-constraints} for a fixed value of $c_d$. To derive these constraints, one should also specify the total ALP width, which greatly affects the constraints derived in the resonant regions. In Fig.~\ref{fig:hadronic-constraints}, we assume for illustration a constant value $\Gamma_a = 10^{-6}$~GeV.~\footnote{See Ref.~\cite{Aloni:2018vki} for a data-driven approach proposed to determine $\Gamma(a\to \mathrm{hadrons})$ for ALP masses in the GeV range.} Smaller (larger) values would imply stronger (weaker) constraints in these regions, without affecting the off-shell ones. We stress that our expressions and experimental/theoretical inputs are given in full generality, so that the correct (resonant) bound could be easily assessed to specific flavor models. For the illustrative case we consider, as shown in Fig.~\ref{fig:hadronic-constraints}, we see that the most stringent constraints indeed come from semileptonic decays such as $K\to \pi \mu e$, $B\to K^{(\ast)} \mu e$ and $B\to K^{(\ast)} \mu \tau$, in the resonant regions. On the other hand, for small masses the most stringent constraints come from the decay $K_L \to \mu e$. Prospects for existing/future experiments are also shown in Fig.~\ref{fig:hadronic-constraints} when available, cf.~Table~\ref{tab:hadronic}. 

\begin{table}[!p]
\centering
  \renewcommand{\arraystretch}{1.5} 
\begin{tabular}{|c|ccc|}
\hline
Decay mode & Exp.~limit & Future prospects & Ref.\\
\hline\hline
$\pi^0\to\mu^\mp e^\pm$ & $3.6\times 10^{-10}$	& -- & \cite{Tanabashi:2018oca} 	\\
$\eta\to\mu^\mp e^\pm$ & $6\times 10^{-6}$	& -- &	\cite{Tanabashi:2018oca} \\
$\eta^\prime\to\mu^\mp e^\pm$ & $4.7\times 10^{-4}$	& -- &	\cite{Tanabashi:2018oca} \\
$K_L\to \mu^\mp e^\pm$	& $4.7 \times 10^{-12}$	& -- &	 \cite{Tanabashi:2018oca} \\
$K^+\to \pi^+\mu^+ e^-$	& $1.3 \times 10^{-11}$	& $\approx 10^{-12}$ &	 \cite{Tanabashi:2018oca} \\
$D\to \mu^\pm e^\mp$	& $1.3\times 10^{-8}$	& -- &	\cite{Tanabashi:2018oca}  \\
$B_d\to \mu^\mp e^\pm$	& $1.0\times 10^{-9}$	& $\approx 2\times 10^{-10}$ &	\cite{Tanabashi:2018oca}  \\
$B_s\to \mu^\mp e^\pm$	& $5.4\times 10^{-9}$	& $\approx 8\times 10^{-10}$ &	\cite{Tanabashi:2018oca}  \\
$B^+\to K^+ \mu^+ e^-$ & $6.4 \times 10^{-9}$	& --	&	\cite{Aaij:2019nmj} \\ 
$B^0\to K^\ast \mu^\mp e^\pm$ &	$1.8 \times 10^{-7}$ &	-- &\cite{Tanabashi:2018oca}	\\ 
\hline \hline
$\tau\to e \pi^0$	& $8.0\times 10^{-8}$	&  $\approx 4\times 10^{-10}$& \cite{Tanabashi:2018oca} 	\\
$\tau\to e \eta$	& $9.2\times 10^{-8}$	& $\approx 9\times 10^{-10}$ &	\cite{Tanabashi:2018oca}  \\
$\tau\to e \eta^\prime$	& $1.6\times 10^{-7}$	& $\approx7 \times 10^{-10}$ &	\cite{Tanabashi:2018oca}  \\
$\tau\to e K_S$	& $2.6\times 10^{-8}$	& $\approx 5\times 10^{-10}$ &	\cite{Tanabashi:2018oca}  \\
$B_d\to \tau^\pm e^\mp$	& $2.8\times 10^{-5}$	& $\approx 1.6\times 10^{-5}$ & \cite{Tanabashi:2018oca} 	\\ 
$B^+\to K^+ \tau^\pm e^\mp$ & $3\times 10^{-5}$	&	$\approx 2.1 \times 10^{-6}$& \cite{Tanabashi:2018oca}	\\ 
\hline\hline
$\tau\to \mu \pi^0$	 & $1.1\times 10^{-7}$	& $\approx 5\times 10^{-10}$ &	\cite{Tanabashi:2018oca}  \\
$\tau\to \mu \eta$	& $6.5\times 10^{-8}$	& $\approx 4\times 10^{-10}$ &	\cite{Tanabashi:2018oca}  \\
$\tau\to \mu \eta^\prime$	& $1.3\times 10^{-7}$	& $\approx 8 \times 10^{-10}$ &	\cite{Tanabashi:2018oca}  \\
$\tau\to \mu K_S$	& $2.3 \times 10^{-8}$	& $\approx 4\times 10^{-10}$ &	\cite{Tanabashi:2018oca}  \\
$B_d\to \tau^\pm \mu^\mp$	& $4.2\times 10^{-5}$	& $\approx 1.3\times 10^{-5}$ &	\cite{Aaij:2019okb}  \\ 
$B_s\to \tau^\pm \mu^\mp$	& $1.4\times 10^{-5}$	& -- &	\cite{Aaij:2019okb} \\ 
$B^+\to K^+ \tau^\pm \mu^\mp$ & $4.8\times 10^{-5}$	& $\approx 3.3 \times 10^{-6}$	& \cite{Tanabashi:2018oca}	\\ \hline
\end{tabular}
\caption{\em \small Most relevant experimental limits on LFV $\tau$ and leptonic meson decays~\cite{Tanabashi:2018oca} and future prospects 
for NA62~\cite{Petrov:2017wza}, LHCb~\cite{Bediaga:2018lhg,Borsato:2018tcz} and Belle-II~\cite{Kou:2018nap,Cerri:2018ypt}.}  
\label{tab:exp}
\end{table}

\newpage

\section{Conclusions}
\label{sec:conclusion}

In this work, we have explored the signatures of axion-like particles (ALPs) in lepton flavor violating (LFV) observables at low energies. 
By using the most general dimension-5 effective Lagrangian, which accounts for the ALP couplings to SM fermions and gauge bosons, 
we have derived complete expressions for the most relevant LFV decays of leptons and hadrons. These general formulae can be applied for any ALP mass, as well as for any choice of ALP couplings, thus generalising and complementing previous results available in the literature~\cite{Heeck:2017xmg,Bauer:2019gfk}.

Purely leptonic observables comprise the decays $\ell_{j} \to \ell_{i} \gamma$, $\ell_{j} \to \ell_{i} \ell_k\ell_k$ and 
$\ell_{j} \to \ell_{i} \gamma\gamma$ as well as $\mu \to e$ conversion in nuclei. We find that, currently,
the most stringent limits concern the $\mu \to e$ transition, and arise from the decay modes $\mu\to 3e$ and $\mu\to e\gamma\gamma$ for $m_a < m_\mu$. For $m_a > m_\mu$, the most stringent constraint arises from $\mu\to e\gamma$, which is going to be superseded in the future by the experimental searches for $\mu \to e$ conversion in nuclei at COMET and Mu2E experiments~\cite{Bartoszek:2014mya,Adamov:2018vin}. Likewise, for the $\tau\to\mu$ and $\tau\to e$ transitions, the three body decays $\tau\to 3\mu$ and $\tau\to e\mu\mu$ set the most stringent 
bounds in the range $ 2m_\mu \lesssim m_a \lesssim m_\tau$, while the radiative decays are the most constraining modes for $m_a > m_\tau$. We have fully explored the complementarity of the different decay modes by showing, in particular, that tree-level contributions to $\ell_j \to \ell_i \ell_k\ell_k$ are not entirely negligible above the threshold of on-shell ALP production, and by estimating these contributions along with their interference with loop-level contributions, see~Fig.~\ref{fig:mue-illustration}. Furthermore, we show that the ALP mass can be inferred from the correlation among the different leptonic processes, as illustrated, for instance, by comparing $\mathcal{B}(\mu \to e \gamma)$ with $\mathcal{B}(\mu \to 3e)$ and $\mathcal{B}(\mu+N \to e+N)$ in Figs.~\ref{fig:mue-illustration} and \ref{fig:mueconv_mueg}, respectively.

Concerning hadronic processes, we have focused on the leptonic decays $P\to \ell_i\ell_j$ and $\tau \to P \ell_j$, and the semileptonic ones $P\to P^\prime (V)\ell_i\ell_j$, where $P^{(\prime)}$ and $V$ stand for pseudoscalar and vector mesons, respectively. Leptonic decays can be induced via ALP couplings to quarks and/or gluons. We find that $\tau$ decays into light unflavored mesons ($\pi^0$, $\eta$ and $\eta^\prime$) are the most sensitive probes of the gluonic coupling, while decays of kaons and $B$-mesons are particularly sensitive to quark-flavor violating ALP interactions. On other hand, semileptonic decays can only probe the vector or axial ALP couplings, depending on the spin of the meson in the final state. For this reason, these processes are very complementary probes of ALP interactions, as we have made explicit for a benchmark scenario in Fig.~\ref{fig:hadronic-constraints}. 

As a by-product of this study, we have also revisited ALP contributions to the anomalous magnetic moment of leptons, 
and we have reassessed the possibility of simultaneously explaining the observed discrepancies in $(g-2)_e$ and $(g-2)_\mu$ 
via ALP contributions. For flavor conserving contributions, we find that very large ALP couplings to electrons and photons are needed. On the other hand, LFV contributions can be especially relevant for $(g-2)_e$ given the large $m_\tau/m_e$ chiral enhancement at the amplitude level, cf.~Eq.~\eqref{eq:gm2_LFV}. This enhancement can be exploited to provide an explanation of these anomalies, by invoking flavor conserving 
contributions of Barr-Zee type for $(g-2)_{\mu}$ and LFV effects for $(g-2)_{e}$, see Fig.~\ref{fig:gm2_LFV}.
The main prediction of this scenario would be large values of $\mathcal{B}(\tau\to e\gamma)$, within reach of Belle-II. 

In summary, ALPs can induce a plethora of low-energy LFV phenomena with specific patterns that we have identified in this paper. 
The ongoing experimental program at present NA62~\cite{Petrov:2017wza}, LHCb~\cite{Bediaga:2018lhg} and Belle-II~\cite{Kou:2018nap} 
experiments, as well as the future ones Mu2E~\cite{Bartoszek:2014mya}, Mu3E~\cite{Blondel:2013ia} and COMET~\cite{Adamov:2018vin}, 
plan to improve the current experimental sensitivities by orders of magnitude, offering many possibilities to discover ALPs indirectly through 
their LFV interactions.

\section*{Acknowledgments}
\label{sec:acknowledgments}

We thank R.~G.~Marcarelli for finding a numerical mistake in the previous version of the manuscript, as well as D.~Be\v{c}irevi\'{c}, F.~Feruglio, J.~Fuentes-Martin, M.~König, F.~Mescia and M.~Passera for useful discussions. This project has received support by the European Union's Horizon 2020 research and innovation programme under the Marie Sklodowska-Curie grant agreement N$^\circ$~674896 (ITN Elusives), by the exchange of researchers project ``The flavor of the invisible universe" funded by the Italian Ministry of Foreign Affairs and International Cooperation (MAECI), and by the Swiss National Science Foundation (SNF) under contract $200021-159720$.


\appendix

\section{Useful formulas}
\label{app:formulas}

\subsection{Loop functions}
\label{app:loop_functions}

In this Appendix, we provide the explicit expressions for the loop functions appearing in Sec.~\ref{sec:lep}:
\begin{align}
g_{\gamma}(x) &= 2 \ln \frac{\Lambda^2}{m_{a}^{2}} -\frac{\log x}{x-1}-(x-1) \log \frac{x}{x-1}-2 \,, \\
g_{1} (x) &= \frac{x-3}{x-1} x^2 \log x +1 -2 x-2 x^{\frac32} \sqrt{x-4}  \log \left(\frac{\sqrt{x}+\sqrt{x-4}}{2}\right) \,, \\
g_{2}(x) &=  1-2 x+2 (x -1) x \log  \frac{x}{x-1} \,, \\
g_{3}(x) &=  \frac{2 x^{2} \log x}{(x-1)^{3}} + \frac{1-3 x}{(x-1)^{2}} \,,\\
g_{4}(x) &= 1 - 2 x + 2 (x-1) x \log \frac{x}{x-1}\,,\\
h_{1}(x) &= 1 +2 x-(x-1) x \log x +  2 x (x  - 3  )  \sqrt{\frac{x }{x-4}} \,\log \left( \frac{\sqrt{x} + \sqrt{x-4}}{2} \right)  \,,\\
h_{2}(x)& =  1 + \frac{x^{2}}{6}   \log x - \frac{ x}{3}-\frac{x+2}{3}
   \sqrt{(x-4) x} \, \log \left( \frac{\sqrt{x} + \sqrt{x-4}}{2} \right)  \,,\\
h_{3}(x) &=  2 x^2 \log \frac{x}{x-1} -1-2 x \,.
\end{align}

\subsection{$\ell_j\to \ell_i \ell_k \ell_k$ kinematic functions}
\label{app:mu3e-kin}

The $\ell_j\to \ell_i \ell_k \ell_k$ phase-space functions that appear in Eq.~\eqref{eq:Gamma-tree} and \eqref{eq:Gamma-int} are given by
\begin{align}
\varphi^{ii}_{0} (x) &= -\frac{11}{4} +4 x- \left[ \frac{x^2}{2} \log \dfrac{2x-1}{x}-1+ 5 x-4x^2\right]\log \dfrac{x-1}{x} \\
&\hspace*{6.em}+ \frac{x^2}{2} \left[\mathrm{Li}_2\left(\dfrac{x-1}{2x-1}\right)-\mathrm{Li}_2\left(\dfrac{x}{2x-1}\right)\right]\,,
\nonumber\\[0.3em]
\varphi^{i \neq k}_{0} (x) & = \left(3 x^2-4 x+1\right) \log \frac{x-1}{x}+3 x-\frac{5}{2}\,, 
\nonumber\\[0.3em]
\varphi_1 (x) &= -3+6x +6(x-1)x \log \dfrac{x-1}{x} \,, \\[0.3em]
\varphi_2 (x) &= 2-2(x-1) \varphi_1(x)\,,
\end{align}
where we remind the reader that $x=m_a^2/m_{\ell_j}$. In the limit of heavy ALP masses, these expressions can be simplified as 
$\varphi^{ii}_{0} (x)=1/(16 x^2) + \mathcal{O}(1/x^3)$, $\varphi^{i \neq k}_{0} (x)= 1/(12 x^2) + \mathcal{O}(1/x^3)$ and  
$\varphi_{1,2} (x)=1/x + \mathcal{O}(1/x^2)$.

\subsection{Form factors for $\ell_{j} \to \ell_{i} \gamma^{\ast}$}
\label{app:offshellff}

In this Appendix we collect the complete one-loop form factors for $\ell_j\to \ell_i \gamma^\ast$ in terms of the Passarino Veltman (PaVe) functions, as defined in the {\sc{Package-X}} documentation~\cite{Patel:2016fam}. The photon 4-momentum is denoted as $q$ and we use the following shorthand notation for the ALP coupling to leptons,
\begin{align}
A_{ab} &\equiv (m_{b} +m_{a})\, a_{ab}^\ell\,,\qquad\qquad
V_{ab} \equiv (m_{b}-m_{a})\, v_{ab}^\ell\,,
\end{align}  
where $m_i \equiv m_{\ell_i}$. The three-point PaVe appearing in our expressions have three possible arguments, which we denote as
\begin{align}
\theta_3&=\lbrace m_j^2,m_i^2,q^2; m_k,m_a,m_k\rbrace\,,\\[0.35em]
\theta_2&=\lbrace m_j^2,q^2,m_i^2; m_a,m_k,m_k\rbrace\,,\\[0.35em]
\theta_1&=\lbrace m_j^2,m_i^2,q^2; 0,m_j,m_a\rbrace\,,\\[0.35em]
\theta_0&=\lbrace m_j^2,m_i^2,q^2; m_a,m_i,0\rbrace\,.
\end{align}
The results provided below can be evaluated by using, for instance, Package-X~\cite{Patel:2016fam}.

\subsubsection{Linear contributions}

For the contributions that are linear on ALP Yukawas, as depicted in the left panel of Fig.~\ref{fig:diagram-mueg}, we obtain that the anapole form-factors read
\begin{align}
\mathcal{F}_{1}(q^{2})  = -q^2\,\frac{e^2\,c_{\gamma \gamma}\, A_{ij} }{8 \pi^2 \Lambda^{2}} (D-3)   &\bigg{\lbrace}(m_j-m_i)\,\Big[\mathcal C_1\left(\theta_0\right)+
\mathcal C_{12}\left(\theta_1\right)+C_{12}\left(\theta_0\right)\Big] \\
   &+m_j\,\Big[\mathcal C_{11}\left(\theta_1\right)+
\mathcal C_{11}\left(\theta_0\right)\Big]\bigg{\rbrace}\,,\nonumber\\[.5em]
\mathcal{G}_{1}(q^{2})  = +q^2\,\frac{e^2\,c_{\gamma \gamma}\, V_{ij} }{8 \pi^2 \Lambda^{2}} (D-3)  &\bigg{\lbrace}(m_j+m_i)\,\Big[\mathcal C_1\left(\theta_0\right)+
\mathcal C_{12}\left(\theta_1\right)+C_{12}\left(\theta_0\right)\Big] \\
   &+m_j\,\Big[\mathcal C_{11}\left(\theta_1\right)+
\mathcal C_{11}\left(\theta_0\right)\Big]\bigg{\rbrace}\,,\nonumber
\end{align}
which vanish at $q^2=0$, as expected by gauge invariance. Similarly, for the dipole form-factors we obtain
\begin{align}
\mathcal{F}_{2}(q^{2}) &= -m_j\,\frac{e^2\,c_{\gamma \gamma}\, A_{ij} }{8 \pi^2 \Lambda^{2}}(D-3)\bigg{\lbrace}(m_j^2-m_i^2)\Big{[}C_1\left(\theta_0\right)+
\mathcal C_{12}\left(\theta_1\right)+C_{12}\left(\theta_0\right)\Big{]}\\
&+m_j(m_j+m_i)\Big{[}\mathcal C_{11}\left(\theta_1\right)+
\mathcal C_{11}\left(\theta_0\right)\Big{]}+2(D-2)\Big{[}\mathcal C_{00}\left(\theta_0\right)+\mathcal C_{00}\left(\theta_1\right)\Big{]} \bigg{\rbrace}\,,\nonumber\\[.5em]
\mathcal{G}_{2}(q^{2}) &= -m_j\,\frac{e^2\,c_{\gamma \gamma}\, V_{ij} }{8 \pi^2 \Lambda^{2}}(D-3)\bigg{\lbrace}(m_j^2-m_i^2)\Big{[}C_1\left(\theta_0\right)+
\mathcal C_{12}\left(\theta_1\right)+C_{12}\left(\theta_0\right)\Big{]}\\
&+m_j(m_j-m_i)\Big{[}\mathcal C_{11}\left(\theta_1\right)+
\mathcal C_{11}\left(\theta_0\right)\Big{]}+2(D-2)\Big{[}\mathcal{C}_{00}\left(\theta_0\right)+\mathcal{C}_{00}\left(\theta_1\right)\Big{]} \bigg{\rbrace}\,.\nonumber
\end{align}
These expressions have been obtained within dimensional regularization, in a scheme where the Levi-Civita symbol is a $D$-dimensional object~\cite{Larin:1993tq}. The scheme choice affects the finite terms for the dipole form-factors since they are UV-sensitive, see discussion in Ref.~\cite{Bauer:2017ris}.

\subsubsection{Quadratic contributions}

For the contributions that are quadratic on ALP Yukawas, as depicted in the right panel of Fig.~\ref{fig:diagram-mueg}, we obtain that the anapole form-factors are given by

\begin{align}
\begin{split}
\mathcal{F}_1(q^2) &= \dfrac{1}{16\pi^2 \Lambda^2}\bigg{[}(A_{ik} A_{kj}+V_{ik}V_{kj})\,\mathcal{F}_1^+(q^2) +(A_{ik} A_{kj}-V_{ik}V_{kj})\,\mathcal{G}_1^-(q^2)\bigg{]}\,,\\
\mathcal{G}_1(q^2) &= \dfrac{1}{16\pi^2 \Lambda^2}\bigg{[}(V_{ik} A_{kj}+A_{ik}V_{kj})\,\mathcal{F}_1^+(q^2) +(V_{ik} A_{kj}-A_{ik}V_{kj})\,\mathcal{G}_1^-(q^2)\bigg{]}\,,
\end{split}
\end{align}

\noindent where
\begin{align}
\mathcal{F}_1^+(q^2)&= m_k\,(m_j+m_i)\, \mathcal{C}_1 \left(\theta_3\right) + \dfrac{m_k}{m_j-m_i} \bigg[\mathcal{B}_0(m_i^2;m_a,m_k)-\mathcal{B}_0(m_j^2;m_a,m_k)\bigg]\,,\\[0.5em]
\mathcal{F}_1^-(q^2)&=2 \, \mathcal{C}_{00}\left(\theta_3\right)+m_j\,(m_j+m_i)\,\mathcal{C}_{11}(\theta_3)+\dfrac{A_0(m_k)-A_0(m_a)}{2\,m_i m_j}-\mathcal{B}_0(q^2;m_k,m_k)\\[0.15em]
&-\dfrac{m_a^2-m_i^2-m_k^2}{2 m_i(m_i-m_j)}\,\mathcal{B}_0(m_i^2;m_a,m_k)-\dfrac{m_a^2-m_j^2-m_k^2}{2 m_j(m_j-m_i)}\,\mathcal{B}_0(m_j^2;m_a,m_k)\nonumber\\[0.4em]
&-(m_a^2-m_j^2-m_k^2)\,\mathcal{C}_0(\theta_2)+(m_j^2-m_i^2)\,\left[\mathcal{C}_{12}(\theta_3)+\mathcal{C}_{2}(\theta_3)\right]+2 m_j^2\,\mathcal{C}_{1}(\theta_3)\,,\nonumber
\end{align}
while $\mathcal{G}_1^{\pm}(q^2)$ is given by $\pm\mathcal{F}_1^{\pm}(q^2)$ with the replacement $m_i\to - m_i$. Lastly, the dipole form-factors read
\begin{align}
\mathcal{F}_{2}(q^2)= \dfrac{m_j}{16\pi^2 \Lambda^2} &\bigg{\lbrace}(A_{ik} A_{kj}-V_{ik}V_{kj}) \Big{[}m_j\, C_{11}\left(\theta_2\right)+m_i\, C_{22}\left(\theta_2\right)+(m_j+m_i)\,C_{12}\left(\theta_2\right)\Big{]}\nonumber\\
&-\Big[(m_k-m_i)
   A_{ik} A_{kj}+(m_k+m_i) V_{ik} V_{kj}\Big]
\mathcal \,C_2 \left(\theta_2\right)\nonumber\\
&-\Big[(m_k-m_j)
   A_{ik} A_{kj}+(m_k+m_j) V_{ik} V_{kj}\Big]
\mathcal \,C_1\left(\theta_2\right)\bigg{\rbrace}\,,\\[0.5em]
\mathcal{G}_{2}(q^2)= \dfrac{m_j}{16\pi^2 \Lambda^2} &\bigg{\lbrace}(V_{ik} A_{kj}-A_{ik}V_{kj}) \Big{[}m_j\, C_{11}\left(\theta_2\right)-m_i\, C_{22}\left(\theta_2\right)+(m_j-m_i)\,C_{12}\left(\theta_2\right)\Big{]}\nonumber\\
&-\Big[(m_k+m_i)
   V_{ik} A_{kj}+(m_k-m_i) A_{ik} V_{kj}\Big]
\mathcal \,C_2 \left(\theta_2\right)\nonumber\\
&-\Big[(m_k-m_j)
   V_{ik} A_{kj}+(m_k+m_j) A_{ik} V_{kj}\Big]
\mathcal \,C_1\left(\theta_2\right)\bigg{\rbrace}\,.
\end{align}

\subsection{Taylor-expanded anapole form factors}
\label{app:taylor}

For $m_a > m_{\ell_j}$, it is a good approximation to Taylor expand the form factors $\mathcal{F}_{1}$ and $\mathcal{G}_{1}$ around $q^2 =0$, cf.~Eq.~\eqref{eq:dF10}. In this Appendix, we provide the explicit expression for the $q^2$-derivative of $\mathcal{F}_{1}(q^2)$ and $\mathcal{G}_{1}(q^2)$ evaluated at $q^2=0$. 

\subsubsection{Linear contributions}

For the contributions illustrated in the right panel of Fig.~\ref{fig:diagram-mueg}, we obtain 
\begin{align}
\begin{split}
\dot{\mathcal{F}}_{1}(0) =  \frac{a_{j i}^\ell\, c_{\gamma \gamma}\,\alpha_\mathrm{em}}{6 \pi \Lambda^{2} }  &\Bigg{[} 6  x_j^2 \, \text{Li}_2 \left(\frac{x_j-1}{x_j}\right)  - \pi ^2 x_j^2 + 3 (x_j + 1)+ 3   (x_j-1) x_j \log \frac{x_j}{x_j-1} \\
&+3 x \frac{2 x_j-1}{x_j-1}  \log x_j \Bigg{]}\,,
\end{split}
\end{align}
while the axial form-factor reads $\dot{\mathcal{G}}_{1}(0)  = -  \dot{\mathcal{F}}_{1}(0)\,\dfrac{v_{j i}^\ell}{a_{j i}^\ell}$\,. 
\subsubsection{Quadratic contributions}

Similarly to the discussion in Sec.~\ref{ssec:dipole} for the dipole operators, the expression for the chirality-conserving form-factors will depend on the mass of the particle running in the loop, as described in the following: 
\begin{itemize}
	\item[•] For $j=k>i$, we obtain that
\begin{align}
\dot{\mathcal{F}}_{1}(0) &=  -\frac{a_{i j}^{\ell} a_{j j}^{\ell}}{16 \pi ^2 \Lambda ^2 }      \left \lbrace  2  x_{j} \frac{2 x_{j}^2-5 x_{j}+2}{x_{j}-1} \log
   x_{j}   + 4 m_j^2 (x_{j}-1)^2 \,\mathcal{C}_0\left(0,0,m_j^2;m_{a},m_j,m_j\right) \right. \nonumber \\
&  \left. \,\,\, + 7-12 x_{j} -  4 (2 x_{j}-1) \sqrt{ (x_{j}-4) x_{j}} \log \left(\frac{\sqrt{x_{j}}+\sqrt{x_{j}-4}}{2}  \right) \right \rbrace\,,\\[0.5em]
\dot{\mathcal{G}}_{1}(0)  &= - \dot{\mathcal{F}}_{1}(0)\, \dfrac{v_{ji}^{\ell}}{a_{ji}^{\ell}}\,.
\end{align}
\item Similarly, for $j>k=i$, we obtain that
\begin{align}
  \dot{\mathcal{F}}_{1}(0)  &= \frac{a_{ii}^{\ell} a_{ij}^{\ell} }{8 \pi ^2 \Lambda ^2} \frac{m_{i}}{m_j}  \left \lbrace \left(4 x_{j}^2-5 x_{j}+1\right) \log
   \frac{x_{j}-1}{x_{j}}+6 x_{j}+(1-2 x_{j}) \log x_{j}-\frac{5}{2}  \right.\nonumber \\
   & \left. +2 m_j^2 x_{j} (1-x_{j}) \, 
   \mathcal{C}_0\left(0,m_i^2,m_j^2;m_i,m_i,m_a\right)+(2 x_{j}-1) \log
   \frac{m_i^2}{m_j^2}\right \rbrace 
\end{align}
\item[•] For $\mu \to e \gamma$, there is an additional contribution from $\tau$-loops (i.e.~$k>j>i$ in our notation), which is given by 
\begin{align}
\dot{\mathcal{F}}_{1}(0)& =  - \frac{a_{\tau e}
  a_{\tau \mu}+v_{\tau e}v_{\tau \mu} }{32 \pi ^2  \Lambda ^2}\,f_1(x_k)\,,\\
  \dot{\mathcal{G}}_{1}(0) &= +  \frac{v_{\tau e}
  a_{\tau \mu}+a_{\tau e}v_{\tau \mu} }{32 \pi ^2 \Lambda ^2}\,f_1(x_k)\,,
\end{align}
where $h_4(x)$ is given by
\begin{align}
f_1(x)&=\frac{-16 x^3+45 x^2+6 (2 x-3) x^2 \log x -36 x+7  }{ 18  (x-1)^4}\,.
\end{align}
\item[•] Lastly, $\tau \to \mu \gamma$ might receive a contribution from electrons running in the loop. In this case,
\begin{align}
\dot{\mathcal{F}}_{1}(0)& = \frac{ a_{ik}^{\ell} a_{kj}^{\ell} + v_{ik}^{\ell} v_{kj}^{\ell} }{{32 \pi ^2
  \Lambda ^2 } } \frac{m_{i}}{m_{j}}  f_{2}(x_{j}) \,,  \\
\dot{\mathcal{G}}_{1}(0)& =   \frac{  v_{ik}^{\ell} a_{kj}^{\ell}+v_{kj}^{\ell} a_{ik}^{\ell}}{32 \pi ^2 \Lambda ^2}  \frac{m_{i}}{m_j} f_{2}(x_{j}) \,,
  \end{align}
with
\begin{align}
f_{2}(x) = &  (2-4 x)  \log \left(x-1\right)+ x (6 -8 x)  \log  \frac{x}{ x-1} + (4 x -2)  \log \frac{m_k^2}{m_j^2} \nonumber \\
 &   +  12 x  -5 - 4 x  \left(x -1 \right) m_{j}^{2} 
   \,\mathcal{C}_0\left(0,0,m_j^2;m_k,m_k,m_a\right)  \,.
 \end{align}

\end{itemize}

\noindent For compactness, we left explicit the dependence on the PaVe function $\mathcal{C}_0$ in the above expressions. This function can be evaluated by using the integral form
\begin{equation}
\mathcal{C}_0(r_{10}^2,r_{12}^2,r_{20}^2;m_0^2,m_1^2,m_2^2) = -\int_0^1 \mathrm{d}x\int_0^{1-y} \mathrm{d}x \dfrac{1}{\Delta(x,y)}\,,
\end{equation}
with
\begin{equation*}
\Delta(x,y)=(1-x-y)\,m_0^2+x\,m_1^2+y\,m_2^2-x(1-x)\, r_{10}^2-y(1-y)\,r_{20}^2+ xy (r_{10}^2+r_{20}^2-r_{21}^2)\,,
\end{equation*}
\noindent or, alternatively, by using {\sc Package-X}~\cite{Patel:2016fam}.


\end{document}